\documentclass[prb,preprint,aps,superscriptaddress,longbibliography]{revtex4-1}

\usepackage{graphicx, epstopdf}
\usepackage{subfigure}
\usepackage{float}
\usepackage{amsmath}
\usepackage{amsfonts}
\usepackage{amssymb}
\usepackage{bm}
\usepackage{subfigure}
\usepackage{caption}
\usepackage{hyperref}
\usepackage{color}
\usepackage{braket}

\begin{document}

\title{Towards Accurate Predictions of Carrier Mobilities and Thermoelectric Performances in 2D Materials}

\author{Yu Wu}
\thanks{These two authors contributed equally}
\affiliation{Key Laboratory of Micro and Nano Photonic Structures (MOE) and Key Laboratory for Information Science of Electromagnetic Waves (MOE) and Department of Optical Science and Engineering, Fudan University, Shanghai 200433, China}
\author{Bowen Hou}
\thanks{These two authors contributed equally}
\affiliation{Key Laboratory of Micro and Nano Photonic Structures (MOE) and Key Laboratory for Information Science of Electromagnetic Waves (MOE) and Department of Optical Science and Engineering, Fudan University, Shanghai 200433, China}
\author{Ying Chen}
\affiliation{Department of Light Sources and Illuminating Engineering, and Academy for Engineering \& Technology, Fudan University, Shanghai, 200433, China}
\author{Jiang Cao}
\affiliation{School of Electronic and Optical Engineering, Nanjing University of Science and Technology, Nanjing 210094, China}
\author{Congcong Ma}
\affiliation{Department of Light Sources and Illuminating Engineering, and Academy for Engineering \& Technology, Fudan University, Shanghai, 200433, China}
\author{Hezhu Shao}
\email{hzshao@wzu.edu.cn}
\affiliation{College of Electrical and Electronic Engineering, Wenzhou University, Wenzhou, 325035, China}
\author{Yiming Zhang}
\affiliation{Key Laboratory of Micro and Nano Photonic Structures (MOE) and Key Laboratory for Information Science of Electromagnetic Waves (MOE) and Department of Optical Science and Engineering, Fudan University, Shanghai 200433, China}
\author{Zixuan Lu}
\affiliation{Department of Light Sources and Illuminating Engineering, and Academy for Engineering \& Technology, Fudan University, Shanghai, 200433, China}
\author{Heyuan Zhu}
\affiliation{Key Laboratory of Micro and Nano Photonic Structures (MOE) and Key Laboratory for Information Science of Electromagnetic Waves (MOE) and Department of Optical Science and Engineering, Fudan University, Shanghai 200433, China}
\author{Zhilai Fang}
\affiliation{Department of Light Sources and Illuminating Engineering, and Academy for Engineering \& Technology, Fudan University, Shanghai, 200433, China}
\author{Rongjun Zhang}
\email{rjzhang@fudan.edu.cn}
\affiliation{Key Laboratory of Micro and Nano Photonic Structures (MOE) and Key Laboratory for Information Science of Electromagnetic Waves (MOE) and Department of Optical Science and Engineering, Fudan University, Shanghai 200433, China}
\author{Hao Zhang}
\email{zhangh@fudan.edu.cn}
\affiliation{Key Laboratory of Micro and Nano Photonic Structures (MOE) and Key Laboratory for Information Science of Electromagnetic Waves (MOE) and Department of Optical Science and Engineering, Fudan University, Shanghai 200433, China}
\affiliation{Nanjing University, National Laboratory of Solid State Microstructure, Nanjing 210093, China}

\date{\today}

\begin{abstract}
The interactions between electrons and lattice vibrational modes play the key role in determining the carrier transport properties, thermoelectric performance and other physical quantities related to phonons in semiconductors. However, for two-dimensional (2D) materials, the widely-used models for carrier transport only consider the interactions between electrons and some specific phonon modes, which usually leads to inaccruate predictions of electrons/phonons transport properties. In this work, comprehensive investigations on full electron-phonon couplings and their influences on carrier mobility and thermoelectric performances of 2D group-IV and V elemental monolayers were performed, and we also analyzed in details the selection rules on electron-phonon couplings using group-theory arguments. Our calculations revealed that, for the cases of shallow dopings where only intravalley scatterings are allowed, the contributions from optical phonon modes are significantly larger than those from acoustic phonon modes in group-IV elemental monolayers, and LA and some specific optical phonon modes contribute significantly to the total intravalley scatterings. When the doping increases and intervalley scatterings are allowed, the intervalley scatterings are much stronger than intravalley scatterings, and ZA/TA/LO phonon modes dominate the intervalley scatterings in monolayer Si, Ge and Sn. The dominant contributions to the total intervalley scatterings are ZA/TO in monolayer P, ZA/TO in monolayer As and TO/LO in monolayer Sb. Based on the thorough investigations on the full electron-phonon couplings, we predict accurately the carrier mobilities and thermoelectric figure of merits in these two elemental crystals, and reveal significant reductions when compared with the calculations based on the widely-used simplified model. Out work not only provides accrurate predictions of carrier transport and thermoelectric performances in these two series of 2D materials by considering full electron-phonon couplings, but also showcases a computational framework for studying related physical quantities in 2D materials.
\end{abstract}

\flushbottom
\maketitle

\thispagestyle{empty}

\section{Introduction}

Two-dimensional (2D) materials offer challenging opportunities for a wide range of applications in nanoscale devices. Among the extraordinary properties in 2D materials, carrier transport properties play the key role in determining the performance in microelectronic, optoelectronic and thermoelectric devices. Extensive studies have been devoted on carrier transport properties in 2D materials, which are dominantly determined by the interaction between electrons and phonons. In the simplest way, the electron-phonon (\textit{el-ph}) interaction can be understood as originated from the electrostatic potential (known as deformation potential) generated by exciting a phonon in the crystal, which in turn affects all carriers directly. Subsequently, the deformation potential approximation (DPA) method based on the intravalley coupling between electrons and long-wavelength longitudial acoustic (LA) phonon modes\cite{Long2009a,Long2011a}, was proposed to estimate or understand the carrier mobilities in many non-polar 2D semiconductors including graphene\cite{Xi2012}, phosphorene\cite{Qiao2014} and transition metal dichalcogenide (TMD)\cite{Du2015}. However, in some polar or highly anisotropic systems, the DPA method usually misestimates the intrinsic carrier mobility, in which the coupling from the longitudinal optical (LO) phonon modes described as the Fr$\ddot{o}$hlich interaction, and the direction-dependent contributions can not be neglected. Furthermore, as previously reported\cite{Nakamura2017,Gunst2016,Gaddemane2016}, the scatterings from flexural ZA phonons contribute dominantly to the total scattering rates in group-IV elemental monolayers (silicene, germanene, and stanene) due to the lack of $\sigma_h$-symmetry in buckled honeycomb structure belonging to $D_{3d}$ point group, which also fails the DPA method. Therefore, depending on the specific materials and the situation of the underlying mechanism, systematically computing energy- and momemtum-dependent \textit{el-ph} interactions is necessary to accurately predict phonon-assisted processes.   

Inspired by the extraordinary properties of graphene, the group-IV elemental materials of silicene (Si), germanene (Ge) and stanene (Sn) as the promising alternatives, have attracted tremendous interests. Studies show that such monolayers all possess Dirac fermions similar to graphene, high mechanical flexibility and high electron mobility, leading to the potential applications in batteries and topological devices\cite{Peng2013,Balendhran2015}. Monolayer Sn was predicted to possess an intrinsic spin-orbit coupling (SOC) gap $\sim100\;\rm{meV}$, which is ideal to realize the quantum spin Hall (QSH) effect at room temperature\cite{Xu2013,Broek2014}. Recently, Jian $et\;al$ successfully synthesized large-scale and high-quality stanene on Sb (111), providing a good platform to study novel phenomena such as QSH effect\cite{Gou2017}.
In contrast to 2D group-IV materials, 2D group-V elemental materials of phosphorene (P), arsenene (As) and antimonene (Sb) with the same buckled honeycomb structure are semiconductors with wide band gaps, probably leading to the applications in transistors with high on/off ratios\cite{Zhu2014} or optoelectric devices operating in the visible range\cite{Zhang2015e}. Recent works reported that, the surface plasmon resonance sensor based on $\beta$-phosphorene/MoS$_2$  heterojunction possesses significantly higher sensitivity compared with conventional or graphene-based SPR sensors\cite{Sharma2018}, and the monolayer arsenene has the potential to serve as anode materials in high-performance Magnesium-ion batteries\cite{Ye2019}. Interestingly, the $\beta$-antimonene undergoes a transformation from a topological semimetal to a topological insulator at 22 bilayer, then to QSH phase at 8 bilayers and finally to a topological trivial semiconductor at 3 or thinner bilayers\cite{Zhang2012}. 

Herein, based on the first-principles method, we systematically investigate the full \textit{el-ph} coupling effects in 2D group-IV and V materials. These two series of 2D materials indeed possess a broad range of carrier mobilities, which may lead to various applications. Compared with group-IV counterparts, group-V materials are more buckled, strengthening the overlap of the $p_z$ orbitals, thus may lead to the enhancement of the interaction between the electrons and ZA phonons. The conduction band and valence band for group-V materials are flatter than group-IV materials, which is beneficial to satisfy the energy conservation condition and further increases the \textit{el-ph} coupling near the Fermi level. To further investigate the roles of different phonon modes, we also derive in details the selection rule for the full \textit{el-ph} scatterings in these two series of materials. Based on the group-theory analysis, we find that, the LA phonons play a dominant role in the hole scattering in arsenene and antimonene, which is in contrast to the previous report that group-V materials have the same scattering mechanism as group-IV materials\cite{Wang2017}. 
For monolayer phosphorene, arsenene and antimonene, despite the different bands and locations of VBM, LA phonon modes with A irreducible representation (irreps) of C$_2$ group near $\Gamma$ point dominate the intravalley scatterings due to the identical symmetry and irreps of the intial and final electronic states. However, phosphorene is the only material with degenerate VBM where the intervalley scattering from one VBM to another is dominated by ZA and TA phonon modes. The VBM of monolayer arsenene and antimonene locates at $\Gamma$ point without degeneracy, hence the intravalley scattering becomes dominant.
Finally, by considering the full \textit{el-ph} scatterings, an accurate prediction of the thermoelectric performances of these two series of materials is achieved.


\section{Results and Discussion}
\subsection{Optimized Configurations and Band Structures}


\begin{figure}
\centering
\includegraphics[width=1\linewidth]{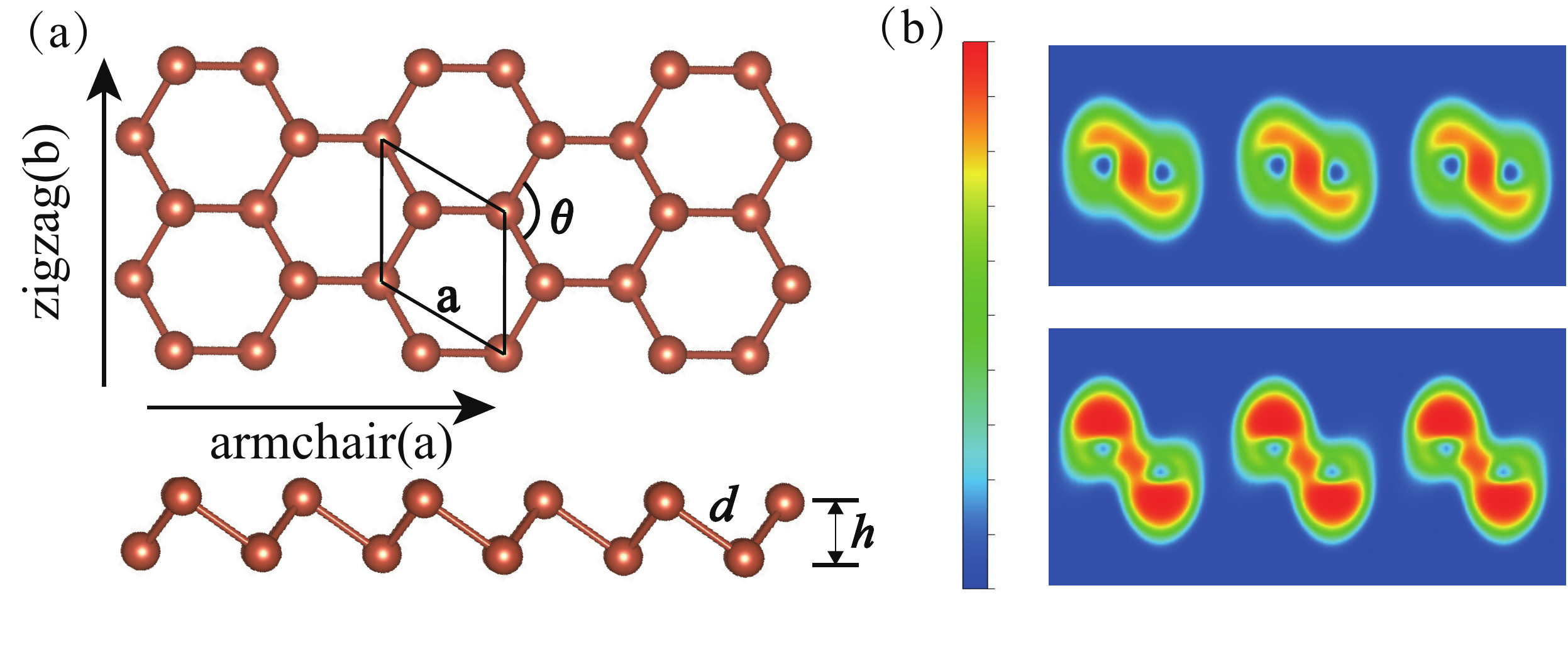}
\caption{(a) Atomic structure of monolayer buckled hexagonal structure.(b) Electron localization function of silicene and phosphorene.}
\label{structure}
\end{figure}

Compared with the planar geometry of graphene, 2D group-IV and V crystals have buckled structure with two sublayers. The top and side views of configurations of the 2D group IV materials (silicene, germanene, stanene) and group V materials (phosphorene, arsenene, antimonene) are shown in Fig.~\ref{structure}(a). The optimized lattice constants $a$, the distance between two sublayers $h$, bond angle $\theta$ and bond length $d$ are shown in Table~S1 which are consistent with previous reports\cite{Mortazavi2016,Kurpas2019,Balendhran2014,Zhang2017}. 

In graphene, the $2s$ orbital is hybridized with $p_x$ and $p_y$ orbitals to form three planar $\sigma$ bonds via $sp^2$ hybridization, and the remaining $p_z$ orbital forms $\pi$ bonding between adjacent C atoms to guarantee the planarity\cite{Cahangirov2009}. However, when the bond length increases, the $p_z$-$p_z$ overlapping decreases and the weaker $\pi$ bonding can not ensure the stability of the planar geometry. By buckling, the planner $sp^2$ bonding is dehybridized to form $sp^3$-like hybridization to enhance the overlapping between $p_z$ orbitals\cite{AkturkE2016}.
Fig.~\ref{structure}(b) shows the calculated electronic localization function (ELF) for silicene and phosphorene. The ELF is a position-dependent function with values ranging from 0 to 1. ELF=1 reflects the maximum probability to find the localized electrons while ELF=0.5 means the electron-gas-like behavior\cite{Silvi1994}. The strength of the $\pi$ bond of 2D group-V materials is much higher than their group-IV counterparts. The one more electron in P, As and Sb increases the degeneracy of electronic states, leading to the unstable state according to the Jahn-Teller theorem, and by a high degree of buckling the degeneracy can be relieved, resulted from the broken symmetry\cite{1937a}. Hence, compared with group-IV materials, the group-V counterparts possess larger buckling heights, smaller lattice constants and bond angles, which together increase the overlap of the $p_z$ orbitals.

In order to further investigate the strength of chemical bonding, we also calculate the bulk ($B_H$) and shear moduli ($G_H$) for group-IV and group-V materials, which determine the ductile-brittle nature of materials by Pugh's ratio $B_H/G_H$\cite{Shao2016}. According to Pugh's rule, a small Pugh's ratio indicates brittle nature of a material. As shown in Table~S1, the Pugh's ratios for phosphorene, arsenene and antimonene are all lower than their buckled group-IV counterparts, indicating that the high degree of buckling in group-V materials strengthens the $\pi$ bonding. The trend of Pugh's ratios $B_H/G_H$(Si)$<$$B_H/G_H$(Ge)$<$$B_H/G_H$(Sn) and $B_H/G_H$(P)$<$$B_H/G_H$(As)$<$$B_H/G_H$(Sb) indicates that increasing bond length $d$ lowers the bonding strength by decreasing orbital overlap for the same group materials, compensating the effect from the increase of buckling.

The calculated band structures of silicene, germanene, and stanene are shown in Fig.~S1(a-c). All these group IV materials are semimetals with a Dirac cone located at the K point, and the SOC effect opens the band gaps of 1.48 meV, 23 meV, 72 meV, respectively. 
Based on the band structures projected by the atomic orbitals, the band structure near the conduction band minimum (CBM, $C_1$) and VBM ($V_1$) are dominantly contributed from $p_z$ orbitals.
However, as shown in Fig.~S1(d-f), as a comparison, the group V materials are all semiconductors with sizable indirect band gaps. 
The calculated band gaps without (with) SOC are 1.94 eV (1.94 eV), 1.59 eV (1.47 eV), 1.26 eV (0.99 eV) for phosphorene, arsenene and antimonene respectively. The CBM ($C_1$) locates close to the middle of the $\Gamma$-M, which is mainly contributed from $p_x$ and $p_z$ orbitals, and the VBM ($V_1$) locates at $\Gamma$ point for arsenene and antimonene which is dominantly contributed from $p_x$
and $p_y$ orbitals, while for phosphorene the VBM locates along $\Gamma$-K direction mainly contributed by $p_z$ orbitals.
The multiple local valleys and peaks, which may play important roles in intervalley \textit{el-ph} scattering, are labeled as $C_2$, $C_3$ and $V_2$, $V_3$,
and their difference from CBM or VBM are shown as $E_{C_2}$, $E_{C_3}$, $E_{V_2}$, $E_{V_3}$ in Table~S1. 
For instance, the energy difference between $V_1$ and $V_2$ valleys ($E_{V_2}$) is $50$ meV in phosphorene.


\begin{table}
\centering
\caption{The energy difference between the valleys (peaks) and CBM (VBM) for all studied group-IV and V materials.}
\begin{tabular}{ccccc}
\hline
 Structure & $E_{C_2}$(eV) & $E_{C_3}$(eV) & $E_{V_2}$(eV) & $E_{V_3}$(eV)\\
\hline
 Si & 1.860 & - & -1.265 & -  \\
 P  & 0.276 & - & -0.049 & -0.162 \\
 Ge & 0.511 & - & -0.427 & -  \\
 As & 0.301 & 0.412 & -0.334 & -0.372 \\
 Sn & 0.201 & - & -0.272 & -  \\
 Sb & 0.252 & 0.293 & -0.437 & -0.466 \\
\hline
\end{tabular}
\label{deltaV}
\end{table}

\subsection{Full \textit{el-ph} scatterings: selection rules and carrier mobilities}

The calculated carrier mobilities at $300$ K along armchair (a) and zigzag (b) directions for these two series of 2D materials, based on the modified DPA theory\cite{Lang2016}, are shown in Table~S1, accompanied with effective mass $m^*$, deformation-potential constant $E_l$, elastic modulus  $C^{2D}$, and the relaxation time $\tau$ defined as $\tau=\mu m^*/e$, which is determined by the interaction between carriers and LA phonons. The deformation-potential constant $E_l$ reflects the coupling strength of electron and long-wavelength LA phonons, defined as $\Delta E_{CBM(VBM)}/(\delta l/l_0)$\cite{Herring1956}, in which $\Delta E_{CBM(VBM)}$ is the shift of the VBM and CBM energy level with respect to the lattice deformation $\delta l$.
The elastic modulus $C^{2D}$ is related to the interatomic forces. Hence, both $E_l$ and $C^{2D}$ are influenced by the overlap of atomic orbitals. The values of $E_l$ and $C^{2D}$ of 2D group-V materials are higher than their group-IV counterparts due to the enhanced overlap of atomic orbitals by buckling. The obtained mobilities for group-IV materials based on the DPA method are comparable to graphene in the order of $10^5\sim10^6\;\rm{cm^{2}V^{-1}s^{-1}}$,  and those of group-V materials are in the range of about $10\sim1000\;\rm{cm^{2}V^{-1}s^{-1}}$. However, as mentioned above, the DPA method may fail for the band-convergence systems such as the group IV and V elemental crystals shown in Fig.~S1, therefore the full \textit{el-ph} scatterings should be investigated to accurately predict the carrier transport properties. According to Eq.~(\ref{miu}), the carrier mobility is mainly determined by the electron velocities and the relaxation times $\tau$. 

\begin{figure*}[ht!]
\centering
\includegraphics[width=1\linewidth]{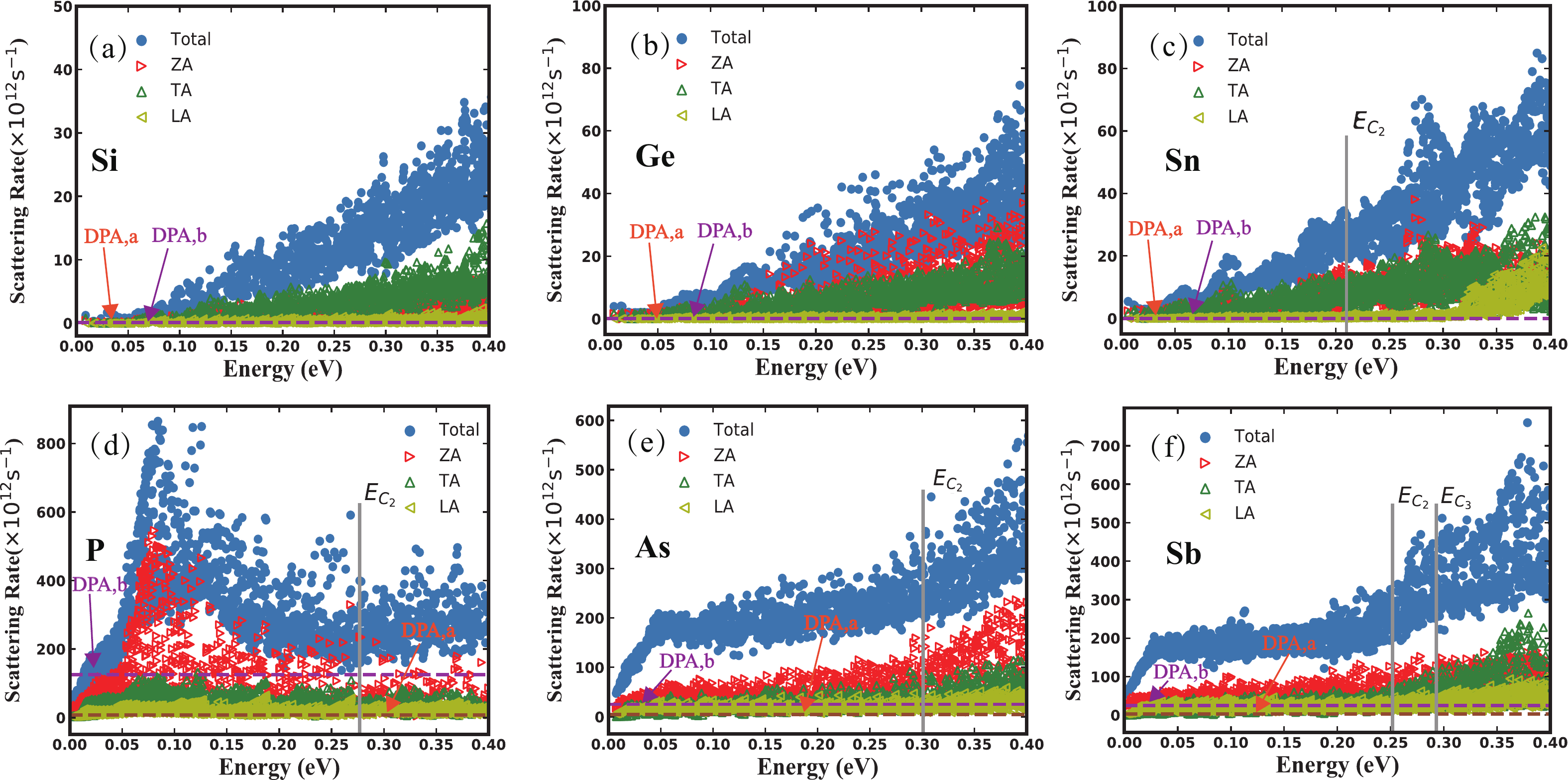}
\caption{The scattering rate of electrons in the $C_1$ valley with energies within $\sim0.4\;\rm{eV}$. The contribution from three acoustic branches and the total scattering rate considering optical branches are demonstrated.}
\label{scattcbm}
\end{figure*}

\begin{figure*}[ht!]
\centering
\includegraphics[width=1\linewidth]{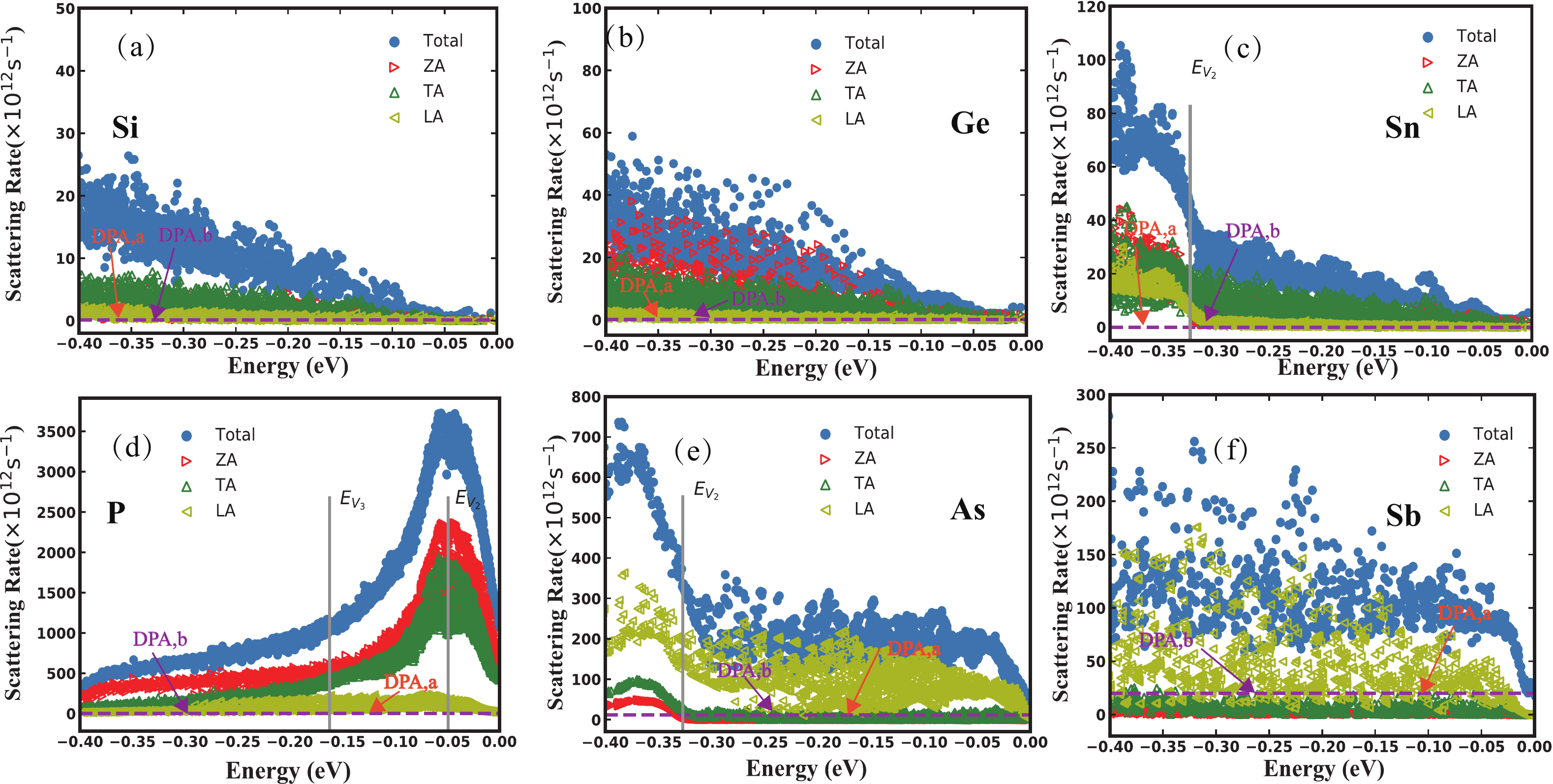}
\caption{The scattering rate of holes in the $V_1$ peak with energies within $\sim0.4\;\rm{eV}$. The case of the $V_2$ peak for stanene,phosphorene and arsenene are also shown in (c-e).}
\label{scattvbm}
\end{figure*}

The calculated \textit{el-ph} scatterings for CBM electrons via acoustic and optical phonon modes for these two series of 2D materials are shown in Fig.~\ref{scattcbm} and S4 respectively, and the calculated \textit{el-ph} scatterings for VBM holes via acoustic and optical phonon modes for these two series of 2D materials are shown in Fig.~\ref{scattvbm} and S5, respectively. All the scattering rates are calculated at 300 K within $\sim0.4\;\rm{eV}$. Obviously, the calculated scattering rates for group-V materials are much larger than their group-IV counterparts, which is probably resulted from two mechanisms: (i) The enhanced $p_z$-$p_z$ orbitals overlapping enlarges the \textit{el-ph} matrix element $g_{mn\nu}\left(\bm{k},\bm{q}\right)$ described in Eq.~(\ref{scarate}) due to the stronger sensitivity to the atomic positions. (ii) The band structures of group-V materials near Fermi level are much flatter than the those of group-IV materials, which is more beneficial to satisfy the energy conservation conditions in the scattering processes. Starting from the CBM, with the increase of electronic energy, the scattering rates increase gradually for group-IV materials while dramatically for group-V materials especially for phosphorene, which possesses a wide and nearly flat region near the $C_1$ valley as shown in Fig.~S1(d), leading to about 4 times larger in the scattering rate at $E=0.07\;\rm{eV}$ than those in arsenene and antimonene. 

As shown in Fig.~\ref{scattcbm} and Tables~S2-S4, for group-IV elemental monolayers (Si, Ge, Sn) with shallow dopings where only intravalley scatterings are allowed, the intravalley scattering rates for CBM electrons via TO and LO phonon modes are significantly larger than those via acoustic phonon mode, and ZO modes also contribute significantly in monolayer Ge. When the doping increases and intervalley scatterings are allowed, the intervalley scatterings rates surpass the intravalley scatterings, and the contributions from ZA, TA and LO phonon modes dominate the intervalley scatterings. However, the situations for group-V elemental monolayers (P, As, Sb) are different, especially for intravalley scatterings. For intravalley scatterings in  monolayer P, the contributions from ZO, LA and TA phonon modes dominate. In monolayer As, the contributions from TO, LO, LA and TA phonon modes dominate, and in monolayer Sb, the contributions from LA, ZO and LO phonon modes dominate. For intervalley scatterings realized by heavy dopings, the contributions from ZA and TO dominate in monolayer P, ZA and TO phonon modes dominate in monolayer As, and TO and LO dominate in monolayer Sb. Furthermore, similar to case in group-IV elemental monolayers, the intervalley scatterings are siginificantly stronger than intravalley scatterings.

\begin{figure}[ht!]
\centering
\includegraphics[width=1\linewidth]{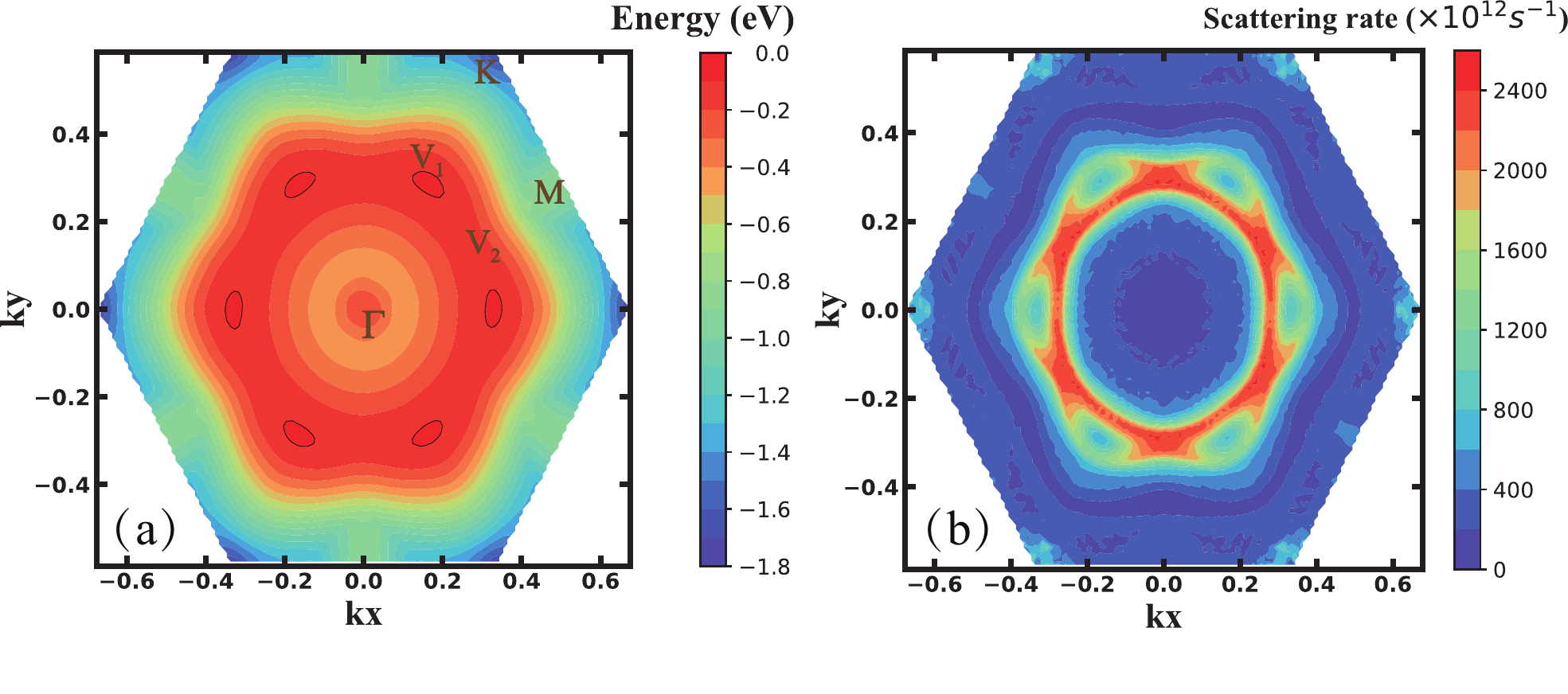}
\caption{Contour map of (a) the valence band and (b) the scattering rate of phosphorene as a function of wave vector $\bm{k}$ at first Brillouin zone.}
\label{Pzone}
\end{figure}

In addition, due to the small phonon energies, for monolayer Sn, As and Sb, since the energy of the $C_1$-valley electrons is well smaller than $E_{C_2}$, only the intravalley and intervalley scatterings between degenerate $C_1$ valleys are possible for small doping. When the energy of $C_1$-pocket electrons is higher than $E_{C_2}$, $C_1$-$C_2$ intervalley scattering is allowed, and the scattering rate increases abruptly as a result. In particular, for monolayer Sb, when the energy of $C_1-$pocket electrons is higher than $E_{C_3}$, both $C_1$-$C_2$ and $C_1$-$C_3$ intervalley scatterings are allowed, leading to a further increase of the scattering rates, as shown in Fig.~\ref{scattcbm}.

The \textit{el-ph} scattering rates of $V_1$-peak holes for these two series of materials at 300 K within $\sim0.4\;\rm{eV}$ are shown in Fig.~\ref{scattvbm}. 
The behaviors of the scattering rates of holes for silicene and germanene are similar to their electron case. 
For stanene, 
as the hole energy descreases to $E_{V_2}$, the scattering rate experiences a sharp rise, implying the activation of $V_1$-$V_2$ intervalley scatterings. 
The intravalley and $V_2$-$V_1$ intervalley scatterings are also allowed for $V_2$-peak holes with energy lower than $E_{V_2}$, and it is found that, the \textit{el-ph} scatterings are dominated by ZA and LA phonons in this case, as shown in Fig.~\ref{scattvbm}(c). 

For the hole scatterings in phosphorene,  due to the flat valence band near Fermi level, also reflected in the effective mass as listed in Table~S1, the scattering rates of $V_1$-peak holes with energy near Fermi level are ultra-high compared with arsenene and antimonene.
When the hole energy is lower than $E_{V_2}$, the $V_1$-$V_2$ intervalley scattering becomes possible and the scattering rates further increase up to $\sim3.7\times10^{15}\;\rm{s^{-1}}$ near $-0.05\;\rm{eV}$. 
%
Fig.~\ref{Pzone}(a) shows the contour map of the valence band in  Brillouin zone for phosphorene. 
It is clear that the energy difference between $V_1$ and $V_2$ peaks is rather small, which is beneficial for satisfying the conservation of energy. 
Fig.~\ref{Pzone}(b) shows the scattering rates of holes from ZA phonons in the valence band in the Brillouin zone, which reveals that, holes at the junction of the two peaks undergo strong scatterings, due to the typical quadratic dispersion of ZA mode and flat electronic dispersion.


In order to further clarify the underlying mechanisms behind electron-phonon interaction, based on the group-theory analysis, the selection rules in terms of the symmetry of systems for the interaction between phonons and electrons are derived.
%
%
%
%
%
The electron-phonon interaction depends on the electronic wave functions of the initial states $\phi^i(\bm{k})$, final states $\phi^f(\bm{k}+\bm{q})$ and the coupled phonon with the eigenvector $\bm{q}$ described by the phonon-induced deformation potential $\delta_{q,\nu} U$\cite{Marlard2009,jiang2005,Castro2007}. 
%
Based on the Fermi's golden rules, in order to calculate the transition probability from initial state $\phi^i(\bm{k})$ to final states $\phi^f(\bm{k}+\bm{q})$ of electron or holes via a specific phonon mode, we only need to obtain the core matrix, which could be expressed as\cite{Chu2014Parity}   

\begin{equation}	
\mathcal{M}(\bm{k},\bm{k'}) = \braket{\phi^i(\bm{k'})|\delta U|\phi^f(\bm{k})}
\end{equation}

where the perturbation potential $\delta U=\sum_{l,a}q_{l,a}\frac{\partial U(r_i)}{\partial q_{l,a}}$, in which $q_{l,a}$ is the ion displacement of the $a$ atom in the $l-$th unit cell, and $r_{i}$ is the electron coordinate. Therefore, the expression for the selection rules for such transition $\mathcal{M}$ based on the group theory is written as,


\begin{equation}
\Gamma^{f} \otimes \Gamma^{ph} \otimes \Gamma^{i} = non\quad null\quad (null)
\label{sim_selection_rule}
\end{equation}

where $\Gamma^{i/f}$ represents the irreps for electron initial/final states, and $\Gamma^{ph}$ denotes the symmetry of the corresponding phonon involved in this process. The involved electrons and phonons must obey the rules of momentum conservation and the compatibility relation among little groups, since they possibly locate at different points in the Brillouin Zone.
\textit{no null/null} means the allowed/forbidden transition channels. Due to the lack of in-plane symmetry $\sigma_h$ in monolayer Si, Ge, Sn of group-IV and monolayer P, As, Sn of group-V, their electronic and phonon symmetry is lowered considerably compared to plane graphene. 

%
%

%
{\color{black}As shown in Figure~S1, for monolayer Si, Ge and Sn, the hole states in the $V_1$ peak along $K$-$M$ and $K$-$\Gamma$ involved in the intrapeak scatterings transform  with the A and B irreps respectively, both belonging to the little group C$_2$. Mathematically, it is allowed for the initial states and final states to belong to the same irreps in the \textit{el-ph} scattering, i.e. $A \rightleftharpoons A$ or $B \rightleftharpoons B$, 
but the contribution of such scattering transition is negligible compared to the transition with different irreps for initial and final states in {\color{black}the real systems we studied here.} Because, as shown in Figure~S1 (a-c), the energy difference of the transition $A \rightleftharpoons B$ is smaller than that for the transition of $A \rightleftharpoons A$ or B $\rightleftharpoons$ B, leading to the larger density of states of the former transition, which is beneficial for the  \textit{el-ph} scattering in real systems. Therefore, for intrapeak scattering in monolayer Si, Ge and Sn, we only need to consider the initial and final states transforming with the A and B irreps, i.e. $A \rightleftharpoons B$ processes.} 
As shown in Tables~S9 and S10, in monolayer Si, Ge and Sn, both ZA and TA modes near $\Gamma$ point transform with the $B$ irreps, while LA modes transform with $A$ irreps.
Hence, restricted by the selection rules shown as Eq.~(\ref{sim_selection_rule}), the hole transition from V$_1^\prime$ to V$_1^{\prime\prime}$ in monolayer Si, Sn and Ge via the out-plane ZA and in-plane TA modes are allowed, due to the \textit{non-null} of $B\otimes B\otimes A$, as shown in Table~S9. The analysis based on the selection rules is in good agreement with the numerical results shown as Fig.~\ref{scattvbm}(a-c) and Tables~S1-S3. 

{\color{black}It should be noticed that, there are some phonon modes near $\Gamma$ point transforming with the irreps belonging to the little group of C$_S$, but it is reasonable to exclude them since the initial and final states of carriers are with C$_2$ symmetry, which are forbidden to be coupled to the phonon modes with C$_S$ group considering the incompatibility relation between $C_2$ and $C_S$\cite{Dresselhaus2008}.}
{\color{black}Furthermore, as shown in Figure~S6, for the interpeak scatterings of holes in group-IV monolayer materials, since the degenerate VBMs are located at six corners in the Brillouin Zone, there are at least three different interpeak-scattering channels, including the transition to the nearest, next-to-nearest and opposite-side K points. However, the interpeak transition to the next-to-nearest VBMs is actually forbidden since the phonon modes involved are with C$_S$ symmetry, which is incompatible to the C$_2$ symmetry. Therefore, only TA modes are allowed restricted by the selection rules and conservation of momentums, as shown in Table~S11.}
The analysis based on the selection rules for holes gives consistent results with previous work\cite{Nakamura2017}, which reported that ZA and TA are dominant in monolayer Si, Ge and Sn. Moreover, as mentioned above, the Dirac points exist in group-IV materials where the CBM and VBM points touch at the $K$ points, therefore, in the same way, the intravalley scatterings of $C_1$-valley electrons scattered via the ZA and TA phonon modes are allowed, while the intervalley scatterings of electrons between degenerate CBMs via TA modes are allowed. The analysis is shown in Tables~S8 and S10.


Similarly, as shown in Figure~S7, the CBMs of group-V materials (monolayer P, As and Sb) locate at the points along $\Gamma$-$M$ with the little group C$_S$, and thus the corresponding irreps of the initial states and final states of CBM are A$^\prime$. 
The acoustic phonon modes of ZA, TA and LA near $\Gamma$ points and along $\Gamma$-$M$ transform with the A$^\prime$, A$^{\prime\prime}$ and A$^\prime$ irreps respectively as shown in Figure~S4 (a-c). Therefore, restricted by the selection rules, both ZA and LA modes are allowed in the intravally scatterings of CBM electrons for monolayer P, As and Sb. The analysis is summarized in Table~S8. For the intervalley scatterings of electrons between degenerate CBMs, the ZA and LA modes are allowed, as shown in Table~S10.

As shown in  Figure~S7(d-f), the VBMs for monolayer As and Sb locate at the $\Gamma$ point, while the VBM for monolayer P locates at the middle point along $\Gamma$-$K$. The initial and final states of VBM holes for monolayer Sb and As near the $\Gamma$ point transform with A irreps belonging to the little group C$_2$. However, despite the different location of the VBM for monolayer P, its little-group symmetry and A irreps for VBM are consistent with Sb and As. As shown in Table~S9, restricted by the selection rules, only LA modes are allowed in the intrapeak scattering for VBM holes in monolayer P, As and Sb, since the LA phonon is the only mode transforming with the A irreps near the $\Gamma$ point, as shown in Fig. S4 (a-c), resulted from the \textit{non null} for the intrapeak transitions. 

Nevertheless, for the interpeak scatterings of the VBM holes for monolayer P, As and Sb, the situation is quite different due to the different locations of VBMs. 
For monolayer As and Sb, since the VBM locates at $\Gamma$ point, which lacks of degeneracy in the Brillouin Zone, the interpeak scatterings between VBMs no longer exist for the V$_1$-peak holes. As a result, only LA mode dominates the \textit{el-ph} scatterings of VBM holes for monolayer As and Sb, which is in good agreement with the mode-resolved scattering rates of $V_1$-peak holes for monolayer As and Sb as shown in Figure~3(e,f) respectively. 
However, special care should be taken for interpeak scatterings of $V_1$-peak holes for monolayer P, since the degeneracy of the VBM points in the Brillouin Zone for monolayer P is six, as shown in Figure~S6 (d).
~Three interpeak transitions to the nearest neighboring local VBMs are possible for the $V_1$-peak holes, and the little groups for the initial and final hole states are $C_2$. However, the phonon involved in the interpeak transition to the secondly nearest neighboring $K$ point transforms with the little group of $C_s$, incompatible to $C_2$, therefore this transition is forbidden. Thus, in monolayer P, the transitions of $V_1$-peak holes to the  nearest and the thirdly nearest neighboring local VBM are allowed, and respectively dominated by the TA and ZA phonon modes. The analysis is shown in Figure~S6 (d) and Table~S11, which is in good agreement with the numerical results shown in Fig.~\ref{scattvbm}(d).

\begin{figure*}
\centering
\includegraphics[width=0.8\linewidth]{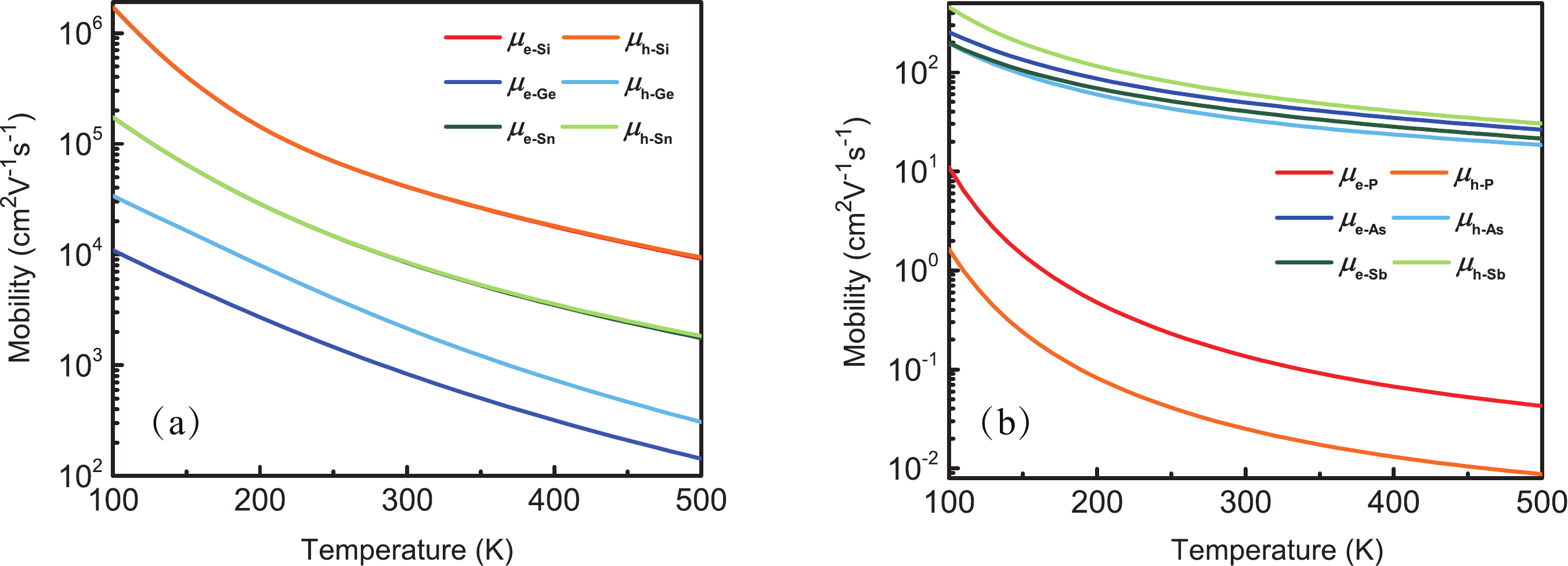}
\caption{The calculated carrier mobility for electrons and holes for (a) group-IV and (b) group-V materials considering full electron-phonon coupling with temperature ranging from 100 K to 500 K.}
\label{mob}
\end{figure*}



Based on the calculated full \textit{el-ph} materix elements, the temperature-dependent and mode-resolved relaxation time and the subsequent carrier mobilites along special directions can be calculated according to Eqs.~(\ref{miu},\ref{scarate}), and the calculated mobilities for these two series of 2D elemental materials are shown in Fig.~\ref{mob} (a) and (b), respectively. The calculated mobilities at 300K, by a comparison with those calculated by the DPA method are shown in Table~S2 as well.
By comparison, three distinctions about the carrier mobilities of 2D group-IV and group-V materials can be identified: \textcolor{black}{(1) Generally, the results by the DPA method are overestimated by $2\sim3$ orders in magnitude due to the only consideration of LA phonons and neglecting the intervalley scattering process; (2) The hole mobilities of group-V materials calculated by these two methods are in roughly good agreement except phosphorene owing to the dominant LA phonons-holes coupling in monolayer $\beta$-Sb and As; (3) The obvious anisotropy in carrier transport for group-V materials obtained by the DPA method no longer exists in the results considering full \textit{el-ph} coupling.}

\subsection{Thermoelectric performance}

The thermoelectric (TE) performance of a material is quantified by a dimensionless figure of merit $zT$$ (zT=S^2\sigma T/(\kappa_{el}+\kappa_l))$, where $S$ represents the Seebeck coefficient, $\sigma$ is the electron conductivity, and $\kappa_{el/l}$ is the electric/lattice thermal conductivity. 
High TE performance needs high $zT$ value of materials with excellent electrical transport properties and poor thermal transport properties simultaneously. However, optimization of $zT$ values in the TE materials is always challenging due to the inter-correlations of the parameters, and optimizing one leads to deteriorating the other. In recent years, several optimization strategies have been proposed to enhance successfully $zT$ values in some traditional TE materials and discover new TE materials, and among them, the strategies of the dimension confinement and band convergence attract interests owning to the successful discovery of layered TE materials with extremely high $zT$ values\cite{Pei2011,lidong2016}. 
Furthermore, considering the 
appropriate band gaps and heavy atoms in arsenene, stanene and antimonene, high $zT$ values can be expected in these two series of 2D elemental materials. 
As reported previously\cite{Chen2017}, antimonene has the $zT$ value of 2.15 at room temperature and can be further enhanced to 2.90 with strain engineering, based on the DPA and constant relaxation time approximation (CRTA) methods.
%
~For arsenene, different constant relaxation times lead to ten times deviation in $zT$ values\cite{Zhang2017}, implying the importance of the accurate calculations of relaxation time in the $zT$ values for a material. 

\begin{figure*}[ht!]
\centering
\includegraphics[width=1\linewidth]{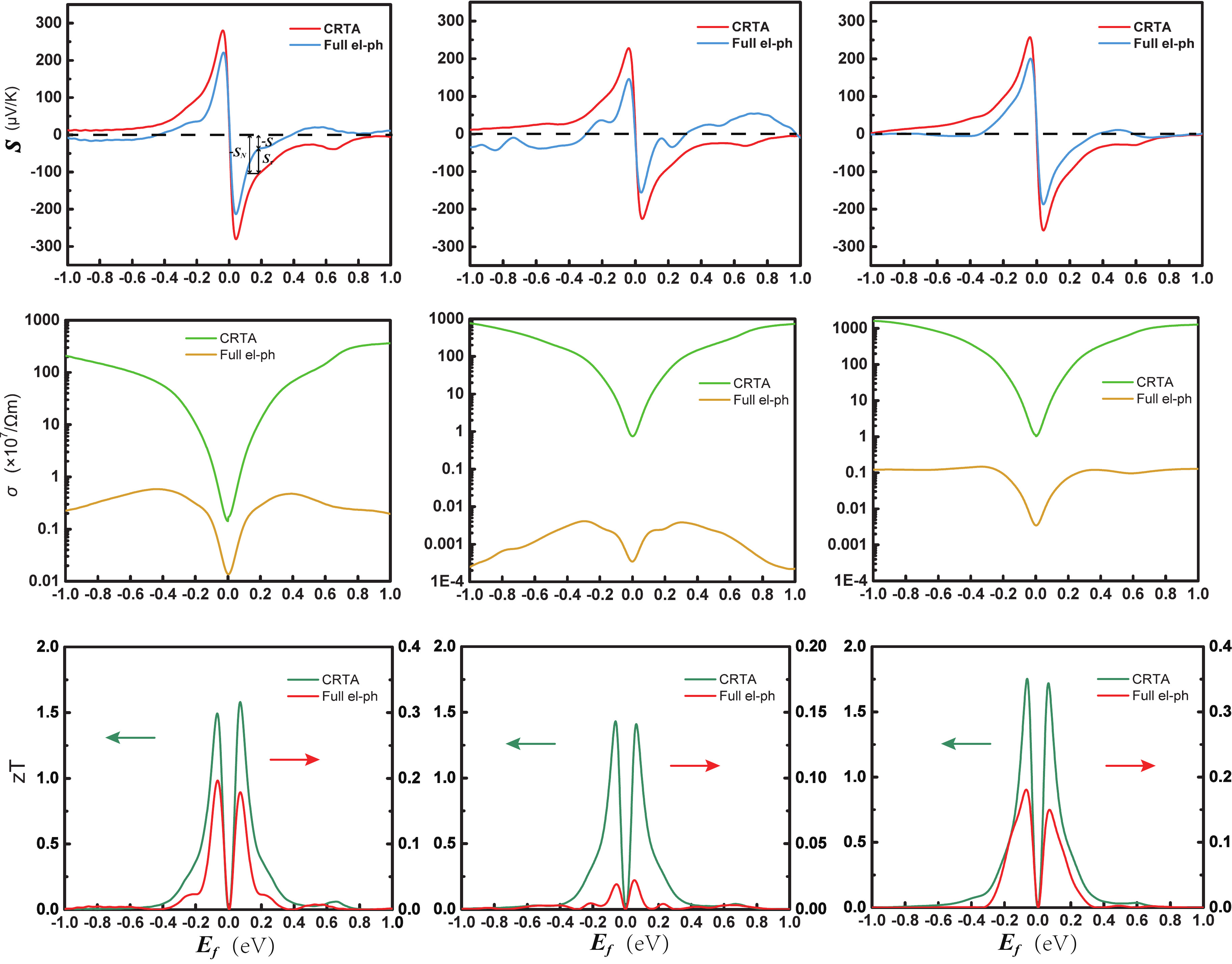}
\caption{The (a) seebeck coefficients (b) electrical conductivity (c) $zT$ value as a function of chemical potential at 300 K for silicene, germanene and stanene.}
\label{TEiv}
\end{figure*}

Here, we also calculate the TE and the related transport properties of these two series of 2D elemental crystals by considering full \textit{el-ph} coupling. Fig.~\ref{TEiv}(a) shows the calculated Seebeck coeffiecient $S$ for the group-IV materials as a function of chemical potential $E_f$ at 300 K using band- and momentum-dependent relaxation times considering full \textit{el-ph} couplings.
 For comparison, the results using the CRTA method are also presented. 
As we know, the behavior of the Seebeck coefficient can be understood by the Mott relation\cite{Sun2015,Wei2015,Liang2016},

\begin{equation}\label{seebeck}
S=-\frac{\pi^{2}}{3} \frac{k_{B}^{2} T}{e}\left[\frac{\partial \ln N(E)}{\partial E}+\frac{\partial \ln \tau(E)}{\partial E}\right]_{E_{f}}
\end{equation}

Where $N(E)$ and $\tau(E)$ are the energy-dependent density of states (DOS) and electronic relaxation time, respectively. 
The Seebeck coefficient can be separated into the band term and the scattering term. A large enhancement of the DOS near the Fermi level leads to a high $S$. 
The scattering term indicates that $S$ is also related to the logarithmic energy derivatives of $\tau(E)$, which is inversely proportional to the scattering rates. In the CRTA method, the scatterings term is equal to zero considering the definition as shown in Eq.~(\ref{seebeck}). 

\begin{figure*}[ht!]
\centering
\includegraphics[width=1\linewidth]{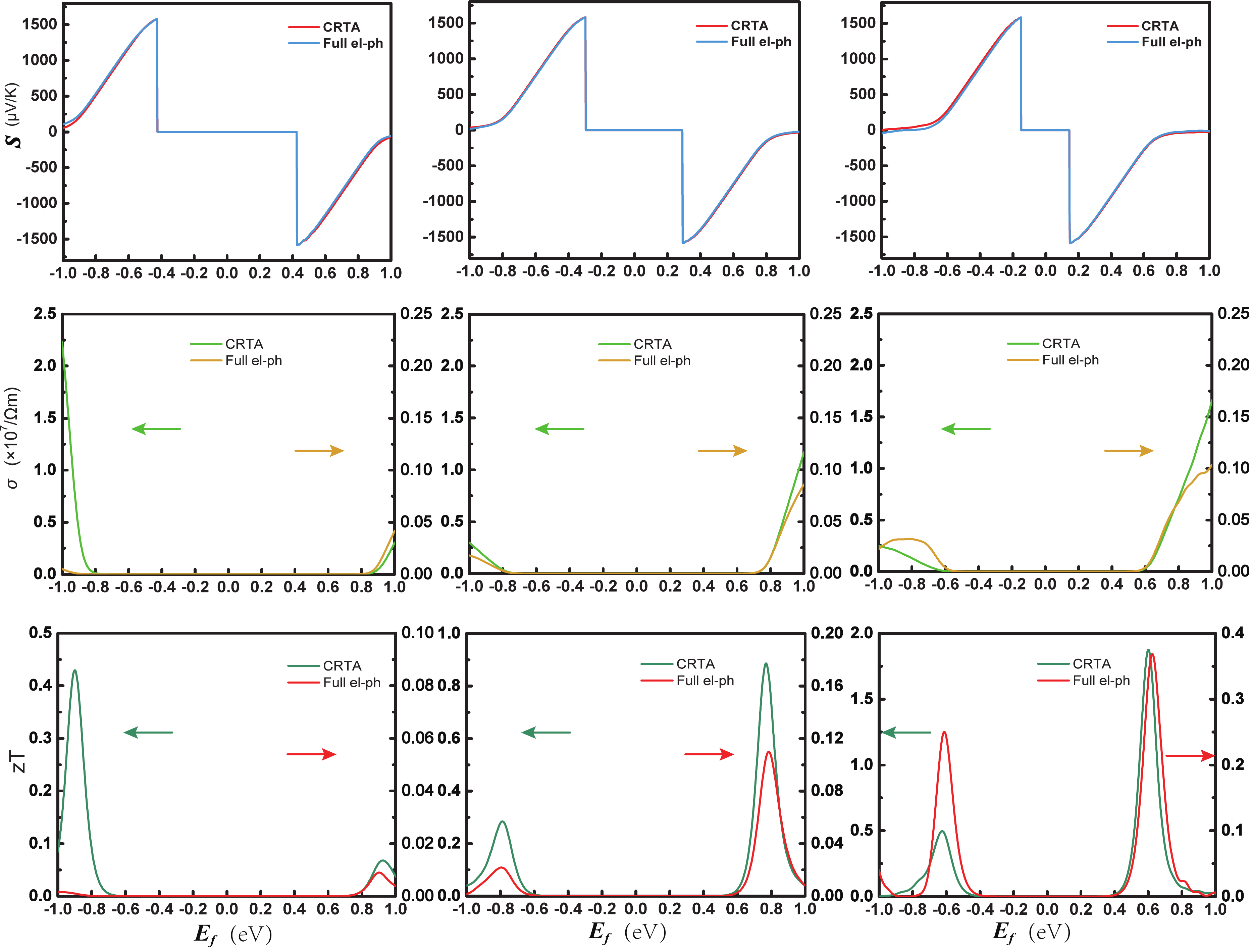}
\caption{The (a) seebeck coefficients (b) electrical conductivity (c) $zT$ value as a function of chemical potential at 300 K for phosphorene, arsenene and antimonene.}
\label{TEv}
\end{figure*}

Considering the n-type systems ($E_f>0$), the scattering rates of electrons gradually increase as shown in Fig.~\ref{scattcbm}(a-c), therefore the scattering term is positive and considerably large, compared with the zero scattering term in the CRTA method. 
Moreover, when the scattering term exceeds the band term, $S$ even undergoes a sign reversal. 
Hence, regarding the relaxation time as a constant for group-IV materials brings significant errors to $S$. 
However, for the group-V materials as shown in Fig.~\ref{TEv}(a), the energy-dependent relaxation time influences little on the Seebeck coefficient resulted from the overwhelming band term in the Seebeck coefficient. 
The maximum Seebeck coefficient of group-V materials is $\sim1500\;\rm{\mu V/K}$, which is nearly 8 times larger than the group-IV materials and many traditional bulk TE materials including Bi$_2$Te$_3$ ($215\;\rm{\mu V/K}$)\cite{Poudel2008}, PbTe ($185\;\rm{\mu V/K}$)\cite{Nielsen2012}, and SnSe ($\sim510\;\rm{\mu V/K}$)\cite{zhao2015}.

For the electrical conductivity $\sigma$ as shown in Fig.~\ref{TEiv}(b) and Fig.~\ref{TEv}(b), near $E_f=0$, $\sigma$ reaches $\sim1\times10^5/\rm{\Omega m}$ ,$\sim1\times10^3/\rm{\Omega m}$, $\sim1\times10^4/\rm{\Omega m}$ for silicene, germanene, and stanene respectively, nearly 1, 4 and 3 orders smaller in magnitude than those calculated using the CRTA method respectively. 
The $\sigma$ for n-type group-V materials are nearly overestimated by an order of magnitude. 
For a $p$-type system, the DPA method gives a better estimation of $\sigma$ for arsenene and antimonene compared with phosphorene. As mentioned above, in phosphorene the intervalley scatterings via ZA phonons are strong which fails the DPA method.


 The total thermal conductivities is the sum of the electronic contribution $\kappa_e$ and the lattice contribution $\kappa_L$, i.e. $\kappa=\kappa_e+\kappa_L$. 
The electronic thermal conductivity $\kappa_e$ exhibits similar trend with $\sigma$ for both DPA and the full \textit{el-ph} scattering methods. 
 According to the Wiedemann-Franz law, $\kappa_e=L\sigma T$, where $L$ is the Lorenz number, increasing $\sigma$ leads to the increase of $\kappa_e$. 
 The lattice thermal conductivities $\kappa_L$ for these two series of 2D elemental crystals are chosen from our previous reports with 28.3 W/mK, 2.4 W/mK, 5.8 W/mK and 106.6 W/mK, 9.0 W/mK, 2.5 W/mK for silicene, germanene, stanene and phosphorene, arsenene, antimonene at 300 K respectively\cite{Peng2017}.

The results of dimensionless figure of merit $zT$ are shown in Fig.~\ref{TEiv}(c) and Fig.~\ref{TEv}(c). 
For group-IV materials, due to the highly overestimated $\sigma$ by the CRTA method, despite the relatively small $S$, the maximum $zT$ value by the CRTA method are in the range of $1.4\sim1.7$, which is much larger than many traditional thermoelectric materials, e.g. Bi$_2$Te$_3$ ($1.2$)\cite{Poudel2008}, PbTe ($0.30$) \cite{Qinyong2013}, SnSe ($0.70$) \cite{Wang2015d}. 
However, the $zT$ values for group-V materials predicted by considering full \textit{el-ph} coupling are only $0.02\sim0.2$, much smaller than those predicted by using the DPA method. 
For example, the p-type phosphorene was reported as a good room-temperature thermoelectric material by the CRTA method with $zT=0.48$, but the $zT$ value decreases to 0.008 when considering full \textit{el-ph} coupling. 
The maximum $zT$ value at 300 K of antimonene reaches $\sim0.4$, which is 5 times smaller than the result of 1.88 predicted by the CRTA method.
%
%
%
~When the temperature rises, the maximum $zT$ value for antimonene at 700 K reaches 1.3 mainly due to the decrease of the $\kappa_L$ ,which is comparable to PbSe (1.1 at 900 K)\cite{Wang2014}, Cu$_2$Se (1.5 at 1000 K)\cite{Liu2012}, PbTe (1.64 at 770 K)\cite{Ahn2013}.

\section{Conclusion}

In summary, using first principles calculations, we systematically investigate the effects of the electron-phonon couplings in 2D honeycomb group-IV and group-V materials with $D_{3d}$ symmetry. High buckling increases the overlap between $p_z$ orbitals and enhances the interaction between carriers and ZA phonons. 
The less dispersive band structure of group-V materials further enhance the electron-phonon coupling due to easier satisfaction of energy conservation condition. {\color{black}We find that the $D_{3d}$ symmetry is not a sufficient condition for ZA phonon scattering to dominate. Based on our results of selection rules, the symmetry of transition process only permits LA phonon to be involved in intravalley scattering within VBM in monolayer phosphorus, arsenene and antimonene. However, TA and ZA modes from intervalley scattering between degenerate VBM is dominant in phosphorus and far beyond LA modes. Yet only LA modes dominate in monolayer arsenene and antimonene since no intervalley scattering between VBM at $\Gamma$ point due to lack of degeneracy.
} 
Moreover, we evaluate the TE performance of all these materials with band and momentum relevant relaxation time indicating that the TE properties of a material can be overestimated with constant relaxation-time approximation from DPA calculation.




\section*{Numerical Methods}
\renewcommand{\theequation}{A\arabic{equation}}
\setcounter{equation}{0}

We carry out density functional theory (DFT) using the QUANTUM ESPRESSO code\cite{Giannozzi2009} with the local density approximation (LDA)\cite{Kresse1996}. Norm-conserving pseudopotentials (NCPP) method with a kinetic energy cutoff of $90$ Ry are used to perform the self-consistent DFT calculation. A vacuum space of 28 \r{A} is set along the perpendicular direction to eliminate the interlayer interacions due to the periodic boundary conditions. For the phonon dispersion, the coarse Monkhorst-Pack $\bm{k}$-mesh and $\bm{q}$-mesh for all the materials discussed are taken as $16\times16\times1$ and $8\times8\times1$ respectively. The Wannier interpolation method is used to generate the ultradense fine grid to describe the processes of electron-phonon scattering accurately in the Wannier90\cite{Mostofi2014} and EPW code\cite{Noffsinger2010,Ponce2016}. Accordingly, much fine $\bm{k}$ and $\bm{q}$ meshes, $240\times240\times1$ and $120\times120\times1$ can be used to calculate the el-ph coupling matrix guaranteeing the numerical convergence of the intrinsic carrier mobility.  
In the Boltzmann transport theory, the electron mobility can be defined as\cite{Xi2014}

\begin{equation}
\mu_{\alpha \beta}=\frac{-e}{n_e\Omega}\sum_{n\in CB}\int\frac{d\bm{k}}{\Omega_{BZ}}  \frac{\partial{f^{0}_{n\bm{k}}}}{\partial{\epsilon_{n\bm{k}}}}v_{n\bm{k},\alpha}v_{n\bm{k},\beta}\tau_{n\bm{k}}
\label{miu}
\end{equation}

Where $n_e$ is the number of electrons, $\Omega$ and $\Omega_{BZ}$ denote the volume of the unit cell and the first Brillouin zone, respectively, $f_{n\bm{k}}^0$ is the Fermi-Dirac distribution, $v_{n\bm{k},\alpha}=h^{-1}\partial{\epsilon_{n\bm{k}}}/\partial{k_{\alpha}}$ is the velocity of the single-particle electron eigenvalue $\epsilon_{n\bm{k}}$. The interaction between electron and phonon lies in the parameter $\tau_{n\bm{k}}$ which is expressed as\cite{Xi2014}

\begin{equation}
\begin{aligned}
\frac{1}{\tau_{n\bm{k}}}=2\rm{Im}\Sigma^{FM}_{n\bm{k}}(\omega)&=\frac{2\pi}{\hbar}\sum_{m\nu}\int\frac{d\bm{q}}{\Omega_{BZ}}\left|g_{mn\nu}\left(\bm{k},\bm{q}\right)\right|^2 \\&\times \left[(1-f^{0}_{m\bm{k}+\bm{q}}+n_{\bm{q}\nu})\delta(\epsilon_{n\bm{k}}-\epsilon_{m\bm{k}+\bm{q}}-\hbar\omega_{\bm{q}\nu})\right.\\&\left. +(f^{0}_{m\bm{k}+\bm{q}}+n_{\bm{q}\nu})\delta(\epsilon_{n\bm{k}}-\epsilon_{m\bm{k}+\bm{q}}+\hbar\omega_{\bm{q}\nu})\right]
\label{scarate}
\end{aligned}
\end{equation}

where the sum is over all the final band index $m$ for electrons and all the phonon mode index $\nu$ and wavevector $\bm{q}$. $\omega_{\bm{q}\nu}$ and $n_{\bm{q}\nu}$ are the frequency and Bose-Einstein distribution of phonons. $\epsilon_{m\bm{k}+\bm{q}}$ and $f^{0}_{m\bm{k}+\bm{q}}$ are the electron eigenvalue and the Fermi-Dirac distribution of the final state with band index $m$ and wavevector $\bm{k'}=\bm{k}+\bm{q}$. The electron-phonon matrix element $g_{mn\nu}\left(\bm{k},\bm{q}\right)$ is calculated using\cite{Baroni2001}

\begin {equation}
g_{mn\nu}(\bm{k},\bm{q})=\left<\phi_{m\bm{k}+\bm{q}}|\Delta_{\bm{q}\nu}V^{KS}|\phi_{n\bm{k}}\right>
\end {equation}

with $\phi_{n\bm{k}}$ and $\phi_{m\bm{k}+\bm{q}}$ being the initial and final electronic Bloch state, respectively. $\Delta_{\bm{q}\nu}V^{KS}$ is the variation of the self-consistent Kohn-Sham (KS) potential experienced by the electrons.

We use BoltzTraP2 code based on the rigid band approach to calculate the transport parameters for arsenene and antimonene monolayers\cite{Madsen2018}. The temperature- and doping-dependent electrical transport properties, including carrier concentrations for holes and electrons $n_{h/e}$, electronic conductivity $\sigma$, electronic thermal conductivities $\kappa_{el}$ and Seebeck coefficient $S$ are computed by solving the semiclassical Boltzmann transport equation (BTE), which can be written as \cite{Madsen2006,Yang2008,Hong2016},

\begin{equation}
n_{\mathrm{h}}(T,\mu)=\frac{2}{\Omega}\iint_{\mathrm{ BZ}}\left[1-f_{0}\left(T,\varepsilon,\mu\right)\right]D(\varepsilon)\mathrm{d}\varepsilon
\label{nh}
\end{equation}

\begin{equation}
n_{\mathrm{e}}(T,\mu)=\frac{2}{\Omega}\iint_{\mathrm{BZ}}f_{ 0}\left(T,\varepsilon,\mu\right)D(\varepsilon)\mathrm{d}\varepsilon.
\label{ne}
\end{equation}

\begin{equation}\label{sigma}
\sigma_{\alpha\beta}(T,\mu)=\frac{1}{\Omega}\int\overline{\sigma}_{\alpha\beta}(\varepsilon)\left[-\frac{ \partial f_{0}(T,\varepsilon,\mu)}{\partial\varepsilon}\right]d\varepsilon,
\end{equation}

\begin{equation}\label{kappa0}
\kappa^{el}_{\alpha\beta}(T,\mu)=\frac{1}{e^2T\Omega}\int\overline{\sigma}_{\alpha\beta}(\varepsilon)(\varepsilon-\mu)^2\left[-\frac{ \partial f_{0}(T,\varepsilon,\mu)}{\partial\varepsilon}\right]d\varepsilon,
\end{equation}

\begin{equation}\label{S}
S_{\alpha\beta}(T,\mu)=\frac{1}{eT\Omega\sigma_{\alpha\beta}(T,\mu)}\int\overline{\sigma}_{\alpha\beta}(\varepsilon)(\varepsilon-\mu )\left[-\frac{\partial f_{0}(T,\varepsilon,\mu)}{\partial\varepsilon}\right]d\varepsilon,
\end{equation}

where $\Omega$ is the unit cell volume, $f_{0}$ is the Fermi-Dirac distribution, $\mu$ is the chemical potential, $D(\varepsilon)$ is the density of states, $\overline{\sigma}_{\alpha\beta}(\varepsilon)$ is the energy dependent conductivity tensor, which can be obtained by $\overline{\sigma}_{\alpha\beta}(\varepsilon)=\frac{1}{N}\sum_{n,\mathbf{k}}\overline{\sigma}_{\alpha\beta}(n,\mathbf{k})\frac{\delta(\varepsilon-\varepsilon_{n,k})}{d\varepsilon}$, in which N is the number of sampled $\mathbf{k}$ points and $\overline{\sigma}_{\alpha\beta}(n,\mathbf{k})$ can be calculated by the formulae based on the kinetic theory, i.e. $\overline{\sigma}_{\alpha\beta}(n,\mathbf{k})=e^2\tau_{n,\mathbf{k}}v_\alpha(i,\mathbf{k})v_\beta(n,\mathbf{k})$. The velocity of carrier $v_{\alpha,\beta}(n,\mathbf{k})$ is defined by $v_i=\frac{1}{\hbar}\frac{\partial \varepsilon_{n
,\mathbf{k}}}{\partial k_i}(i=\alpha,\beta)$. In this code, the electron-phonon relaxation time is treated as both energy and direction dependent from Eq.(\ref{scarate}). 

\section*{Acknowledgement}

This work is supported by the National Natural Science Foundation of China under Grants No. 11374063 and 11404348.







\begin{thebibliography}{61}%
\makeatletter
\providecommand \@ifxundefined [1]{%
 \@ifx{#1\undefined}
}%
\providecommand \@ifnum [1]{%
 \ifnum #1\expandafter \@firstoftwo
 \else \expandafter \@secondoftwo
 \fi
}%
\providecommand \@ifx [1]{%
 \ifx #1\expandafter \@firstoftwo
 \else \expandafter \@secondoftwo
 \fi
}%
\providecommand \natexlab [1]{#1}%
\providecommand \enquote  [1]{``#1''}%
\providecommand \bibnamefont  [1]{#1}%
\providecommand \bibfnamefont [1]{#1}%
\providecommand \citenamefont [1]{#1}%
\providecommand \href@noop [0]{\@secondoftwo}%
\providecommand \href [0]{\begingroup \@sanitize@url \@href}%
\providecommand \@href[1]{\@@startlink{#1}\@@href}%
\providecommand \@@href[1]{\endgroup#1\@@endlink}%
\providecommand \@sanitize@url [0]{\catcode `\\12\catcode `\$12\catcode
  `\&12\catcode `\#12\catcode `\^12\catcode `\_12\catcode `\%12\relax}%
\providecommand \@@startlink[1]{}%
\providecommand \@@endlink[0]{}%
\providecommand \url  [0]{\begingroup\@sanitize@url \@url }%
\providecommand \@url [1]{\endgroup\@href {#1}{\urlprefix }}%
\providecommand \urlprefix  [0]{URL }%
\providecommand \Eprint [0]{\href }%
\providecommand \doibase [0]{http://dx.doi.org/}%
\providecommand \selectlanguage [0]{\@gobble}%
\providecommand \bibinfo  [0]{\@secondoftwo}%
\providecommand \bibfield  [0]{\@secondoftwo}%
\providecommand \translation [1]{[#1]}%
\providecommand \BibitemOpen [0]{}%
\providecommand \bibitemStop [0]{}%
\providecommand \bibitemNoStop [0]{.\EOS\space}%
\providecommand \EOS [0]{\spacefactor3000\relax}%
\providecommand \BibitemShut  [1]{\csname bibitem#1\endcsname}%
\let\auto@bib@innerbib\@empty
\bibitem [{\citenamefont {Long}\ \emph {et~al.}(2009)\citenamefont {Long},
  \citenamefont {Tang}, \citenamefont {Wang}, \citenamefont {Wang},\ and\
  \citenamefont {Shuai}}]{Long2009a}%
  \BibitemOpen
  \bibfield  {author} {\bibinfo {author} {\bibfnamefont {Meng-Qiu}\
  \bibnamefont {Long}}, \bibinfo {author} {\bibfnamefont {Ling}\ \bibnamefont
  {Tang}}, \bibinfo {author} {\bibfnamefont {Dong}\ \bibnamefont {Wang}},
  \bibinfo {author} {\bibfnamefont {Linjun}\ \bibnamefont {Wang}}, \ and\
  \bibinfo {author} {\bibfnamefont {Zhigang}\ \bibnamefont {Shuai}},\
  }\bibfield  {title} {\enquote {\bibinfo {title} {Theoretical predictions of
  size-dependent carrier mobility and polarity in graphene},}\ }\href {\doibase
  10.1021/ja907528a} {\bibfield  {journal} {\bibinfo  {journal} {Journal of the
  American Chemical Society}\ }\textbf {\bibinfo {volume} {131}},\ \bibinfo
  {pages} {17728--17729} (\bibinfo {year} {2009})}\BibitemShut {NoStop}%
\bibitem [{\citenamefont {Long}\ \emph {et~al.}(2011)\citenamefont {Long},
  \citenamefont {Tang}, \citenamefont {Wang}, \citenamefont {Li},\ and\
  \citenamefont {Shuai}}]{Long2011a}%
  \BibitemOpen
  \bibfield  {author} {\bibinfo {author} {\bibfnamefont {Mengqiu}\ \bibnamefont
  {Long}}, \bibinfo {author} {\bibfnamefont {Ling}\ \bibnamefont {Tang}},
  \bibinfo {author} {\bibfnamefont {Dong}\ \bibnamefont {Wang}}, \bibinfo
  {author} {\bibfnamefont {Yuliang}\ \bibnamefont {Li}}, \ and\ \bibinfo
  {author} {\bibfnamefont {Zhigang}\ \bibnamefont {Shuai}},\ }\bibfield
  {title} {\enquote {\bibinfo {title} {Electronic structure and carrier
  mobility in graphdiyne sheet and nanoribbons: Theoretical predictions},}\
  }\href {\doibase 10.1021/nn102472s} {\bibfield  {journal} {\bibinfo
  {journal} {{ACS} Nano}\ }\textbf {\bibinfo {volume} {5}},\ \bibinfo {pages}
  {2593--2600} (\bibinfo {year} {2011})}\BibitemShut {NoStop}%
\bibitem [{\citenamefont {Xi}\ \emph {et~al.}(2012)\citenamefont {Xi},
  \citenamefont {Long}, \citenamefont {Tang}, \citenamefont {Wang},\ and\
  \citenamefont {Shuai}}]{Xi2012}%
  \BibitemOpen
  \bibfield  {author} {\bibinfo {author} {\bibfnamefont {Jinyang}\ \bibnamefont
  {Xi}}, \bibinfo {author} {\bibfnamefont {Mengqiu}\ \bibnamefont {Long}},
  \bibinfo {author} {\bibfnamefont {Ling}\ \bibnamefont {Tang}}, \bibinfo
  {author} {\bibfnamefont {Dong}\ \bibnamefont {Wang}}, \ and\ \bibinfo
  {author} {\bibfnamefont {Zhigang}\ \bibnamefont {Shuai}},\ }\bibfield
  {title} {\enquote {\bibinfo {title} {First-principles prediction of charge
  mobility in carbon and organic nanomaterials},}\ }\href {\doibase
  10.1039/c2nr30585b} {\bibfield  {journal} {\bibinfo  {journal} {Nanoscale}\
  }\textbf {\bibinfo {volume} {4}},\ \bibinfo {pages} {4348} (\bibinfo {year}
  {2012})}\BibitemShut {NoStop}%
\bibitem [{\citenamefont {Qiao}\ \emph {et~al.}({2014})\citenamefont {Qiao},
  \citenamefont {Kong}, \citenamefont {Hu}, \citenamefont {Yang},\ and\
  \citenamefont {Ji}}]{Qiao2014}%
  \BibitemOpen
  \bibfield  {author} {\bibinfo {author} {\bibfnamefont {Jingsi}\ \bibnamefont
  {Qiao}}, \bibinfo {author} {\bibfnamefont {Xianghua}\ \bibnamefont {Kong}},
  \bibinfo {author} {\bibfnamefont {Zhi-Xin}\ \bibnamefont {Hu}}, \bibinfo
  {author} {\bibfnamefont {Feng}\ \bibnamefont {Yang}}, \ and\ \bibinfo
  {author} {\bibfnamefont {Wei}\ \bibnamefont {Ji}},\ }\bibfield  {title}
  {\enquote {\bibinfo {title} {{High-mobility transport anisotropy and linear
  dichroism in few-layer black phosphorus}},}\ }\href {\doibase
  {10.1038/ncomms5475}} {\bibfield  {journal} {\bibinfo  {journal} {{NATURE
  COMMUNICATIONS}}\ }\textbf {\bibinfo {volume} {{5}}},\ \bibinfo {pages}
  {4475} (\bibinfo {year} {{2014}})}\BibitemShut {NoStop}%
\bibitem [{\citenamefont {Du}\ \emph {et~al.}(2015)\citenamefont {Du},
  \citenamefont {Liu}, \citenamefont {Xu}, \citenamefont {Sheng}, \citenamefont
  {Yin}, \citenamefont {Duan},\ and\ \citenamefont {Wan}}]{Du2015}%
  \BibitemOpen
  \bibfield  {author} {\bibinfo {author} {\bibfnamefont {Yongping}\
  \bibnamefont {Du}}, \bibinfo {author} {\bibfnamefont {Huimei}\ \bibnamefont
  {Liu}}, \bibinfo {author} {\bibfnamefont {Bo}~\bibnamefont {Xu}}, \bibinfo
  {author} {\bibfnamefont {Li}~\bibnamefont {Sheng}}, \bibinfo {author}
  {\bibfnamefont {Jiang}\ \bibnamefont {Yin}}, \bibinfo {author} {\bibfnamefont
  {Chun-Gang}\ \bibnamefont {Duan}}, \ and\ \bibinfo {author} {\bibfnamefont
  {Xiangang}\ \bibnamefont {Wan}},\ }\bibfield  {title} {\enquote {\bibinfo
  {title} {Unexpected magnetic semiconductor behavior in zigzag phosphorene
  nanoribbons driven by half-filled one dimensional band},}\ }\href {\doibase
  10.1038/srep08921} {\bibfield  {journal} {\bibinfo  {journal} {Scientific
  Reports}\ }\textbf {\bibinfo {volume} {5}},\ \bibinfo {pages} {8921}
  (\bibinfo {year} {2015})}\BibitemShut {NoStop}%
\bibitem [{\citenamefont {Nakamura}\ \emph {et~al.}(2017)\citenamefont
  {Nakamura}, \citenamefont {Zhao}, \citenamefont {Xi}, \citenamefont {Shi},
  \citenamefont {Wang},\ and\ \citenamefont {Shuai}}]{Nakamura2017}%
  \BibitemOpen
  \bibfield  {author} {\bibinfo {author} {\bibfnamefont {Yuma}\ \bibnamefont
  {Nakamura}}, \bibinfo {author} {\bibfnamefont {Tianqi}\ \bibnamefont {Zhao}},
  \bibinfo {author} {\bibfnamefont {Jinyang}\ \bibnamefont {Xi}}, \bibinfo
  {author} {\bibfnamefont {Wen}\ \bibnamefont {Shi}}, \bibinfo {author}
  {\bibfnamefont {Dong}\ \bibnamefont {Wang}}, \ and\ \bibinfo {author}
  {\bibfnamefont {Zhigang}\ \bibnamefont {Shuai}},\ }\bibfield  {title}
  {\enquote {\bibinfo {title} {Intrinsic charge transport in stanene: Roles of
  bucklings and electron-phonon couplings},}\ }\href {\doibase
  10.1002/aelm.201700143} {\bibfield  {journal} {\bibinfo  {journal} {Advanced
  Electronic Materials}\ }\textbf {\bibinfo {volume} {3}},\ \bibinfo {pages}
  {1700143} (\bibinfo {year} {2017})}\BibitemShut {NoStop}%
\bibitem [{\citenamefont {Gunst}\ \emph {et~al.}(2016)\citenamefont {Gunst},
  \citenamefont {Markussen}, \citenamefont {Stokbro},\ and\ \citenamefont
  {Brandbyge}}]{Gunst2016}%
  \BibitemOpen
  \bibfield  {author} {\bibinfo {author} {\bibfnamefont {Tue}\ \bibnamefont
  {Gunst}}, \bibinfo {author} {\bibfnamefont {Troels}\ \bibnamefont
  {Markussen}}, \bibinfo {author} {\bibfnamefont {Kurt}\ \bibnamefont
  {Stokbro}}, \ and\ \bibinfo {author} {\bibfnamefont {Mads}\ \bibnamefont
  {Brandbyge}},\ }\bibfield  {title} {\enquote {\bibinfo {title}
  {First-principles method for electron-phonon coupling and electron mobility:
  Applications to two-dimensional materials},}\ }\href {\doibase
  10.1103/physrevb.93.035414} {\bibfield  {journal} {\bibinfo  {journal}
  {Physical Review B}\ }\textbf {\bibinfo {volume} {93}},\ \bibinfo {pages}
  {035414} (\bibinfo {year} {2016})}\BibitemShut {NoStop}%
\bibitem [{\citenamefont {Gaddemane}\ \emph {et~al.}(2016)\citenamefont
  {Gaddemane}, \citenamefont {Vandenberghe},\ and\ \citenamefont
  {Fischetti}}]{Gaddemane2016}%
  \BibitemOpen
  \bibfield  {author} {\bibinfo {author} {\bibfnamefont {Gautam}\ \bibnamefont
  {Gaddemane}}, \bibinfo {author} {\bibfnamefont {William~G.}\ \bibnamefont
  {Vandenberghe}}, \ and\ \bibinfo {author} {\bibfnamefont {Massimo~V.}\
  \bibnamefont {Fischetti}},\ }\bibfield  {title} {\enquote {\bibinfo {title}
  {Theoretical study of electron transport in silicene and germanene using
  full-band monte carlo simulations},}\ }in\ \href {\doibase
  10.1109/sispad.2016.7605219} {\emph {\bibinfo {booktitle} {2016 International
  Conference on Simulation of Semiconductor Processes and Devices
  ({SISPAD})}}}\ (\bibinfo  {publisher} {{IEEE}},\ \bibinfo {year}
  {2016})\BibitemShut {NoStop}%
\bibitem [{\citenamefont {Peng}\ \emph {et~al.}(2013)\citenamefont {Peng},
  \citenamefont {Wen},\ and\ \citenamefont {De}}]{Peng2013}%
  \BibitemOpen
  \bibfield  {author} {\bibinfo {author} {\bibfnamefont {Qing}\ \bibnamefont
  {Peng}}, \bibinfo {author} {\bibfnamefont {Xiaodong}\ \bibnamefont {Wen}}, \
  and\ \bibinfo {author} {\bibfnamefont {Suvranu}\ \bibnamefont {De}},\
  }\bibfield  {title} {\enquote {\bibinfo {title} {Mechanical stabilities of
  silicene},}\ }\href {\doibase 10.1039/c3ra41347k} {\bibfield  {journal}
  {\bibinfo  {journal} {{RSC} Advances}\ }\textbf {\bibinfo {volume} {3}},\
  \bibinfo {pages} {13772} (\bibinfo {year} {2013})}\BibitemShut {NoStop}%
\bibitem [{\citenamefont {Balendhran}\ \emph {et~al.}(2015)\citenamefont
  {Balendhran}, \citenamefont {Walia}, \citenamefont {Nili}, \citenamefont
  {Sriram},\ and\ \citenamefont {Bhaskaran}}]{Balendhran2015}%
  \BibitemOpen
  \bibfield  {author} {\bibinfo {author} {\bibfnamefont {Sivacarendran}\
  \bibnamefont {Balendhran}}, \bibinfo {author} {\bibfnamefont {Sumeet}\
  \bibnamefont {Walia}}, \bibinfo {author} {\bibfnamefont {Hussein}\
  \bibnamefont {Nili}}, \bibinfo {author} {\bibfnamefont {Sharath}\
  \bibnamefont {Sriram}}, \ and\ \bibinfo {author} {\bibfnamefont {Madhu}\
  \bibnamefont {Bhaskaran}},\ }\bibfield  {title} {\enquote {\bibinfo {title}
  {Elemental analogues of graphene: Silicene, germanene, stanene, and
  phosphorene},}\ }\href {\doibase 10.1002/smll.201402041} {\bibfield
  {journal} {\bibinfo  {journal} {Small}\ }\textbf {\bibinfo {volume} {11}},\
  \bibinfo {pages} {640--652} (\bibinfo {year} {2015})},\ \Eprint
  {http://arxiv.org/abs/https://onlinelibrary.wiley.com/doi/pdf/10.1002/smll.201402041}
  {https://onlinelibrary.wiley.com/doi/pdf/10.1002/smll.201402041} \BibitemShut
  {NoStop}%
\bibitem [{\citenamefont {Xu}\ \emph {et~al.}(2013)\citenamefont {Xu},
  \citenamefont {Yan}, \citenamefont {Zhang}, \citenamefont {Wang},
  \citenamefont {Xu}, \citenamefont {Tang}, \citenamefont {Duan},\ and\
  \citenamefont {Zhang}}]{Xu2013}%
  \BibitemOpen
  \bibfield  {author} {\bibinfo {author} {\bibfnamefont {Yong}\ \bibnamefont
  {Xu}}, \bibinfo {author} {\bibfnamefont {Binghai}\ \bibnamefont {Yan}},
  \bibinfo {author} {\bibfnamefont {Hai-Jun}\ \bibnamefont {Zhang}}, \bibinfo
  {author} {\bibfnamefont {Jing}\ \bibnamefont {Wang}}, \bibinfo {author}
  {\bibfnamefont {Gang}\ \bibnamefont {Xu}}, \bibinfo {author} {\bibfnamefont
  {Peizhe}\ \bibnamefont {Tang}}, \bibinfo {author} {\bibfnamefont {Wenhui}\
  \bibnamefont {Duan}}, \ and\ \bibinfo {author} {\bibfnamefont {Shou-Cheng}\
  \bibnamefont {Zhang}},\ }\bibfield  {title} {\enquote {\bibinfo {title}
  {Large-gap quantum spin hall insulators in tin films},}\ }\href {\doibase
  10.1103/physrevlett.111.136804} {\bibfield  {journal} {\bibinfo  {journal}
  {Physical Review Letters}\ }\textbf {\bibinfo {volume} {111}},\ \bibinfo
  {pages} {136804} (\bibinfo {year} {2013})}\BibitemShut {NoStop}%
\bibitem [{\citenamefont {van~den Broek}\ \emph {et~al.}(2014)\citenamefont
  {van~den Broek}, \citenamefont {Houssa}, \citenamefont {Scalise},
  \citenamefont {Pourtois}, \citenamefont {Afanas`ev},\ and\ \citenamefont
  {Stesmans}}]{Broek2014}%
  \BibitemOpen
  \bibfield  {author} {\bibinfo {author} {\bibfnamefont {B}~\bibnamefont
  {van~den Broek}}, \bibinfo {author} {\bibfnamefont {M}~\bibnamefont
  {Houssa}}, \bibinfo {author} {\bibfnamefont {E}~\bibnamefont {Scalise}},
  \bibinfo {author} {\bibfnamefont {G}~\bibnamefont {Pourtois}}, \bibinfo
  {author} {\bibfnamefont {V~V}\ \bibnamefont {Afanas`ev}}, \ and\ \bibinfo
  {author} {\bibfnamefont {A}~\bibnamefont {Stesmans}},\ }\bibfield  {title}
  {\enquote {\bibinfo {title} {Two-dimensional hexagonal tin:ab initiogeometry,
  stability, electronic structure and functionalization},}\ }\href {\doibase
  10.1088/2053-1583/1/2/021004} {\bibfield  {journal} {\bibinfo  {journal} {2D
  Materials}\ }\textbf {\bibinfo {volume} {1}},\ \bibinfo {pages} {021004}
  (\bibinfo {year} {2014})}\BibitemShut {NoStop}%
\bibitem [{\citenamefont {Gou}\ \emph {et~al.}(2017)\citenamefont {Gou},
  \citenamefont {Kong}, \citenamefont {Li}, \citenamefont {Zhong},
  \citenamefont {Li}, \citenamefont {Cheng}, \citenamefont {Chen},\ and\
  \citenamefont {Wu}}]{Gou2017}%
  \BibitemOpen
  \bibfield  {author} {\bibinfo {author} {\bibfnamefont {Jian}\ \bibnamefont
  {Gou}}, \bibinfo {author} {\bibfnamefont {Longjuan}\ \bibnamefont {Kong}},
  \bibinfo {author} {\bibfnamefont {Hui}\ \bibnamefont {Li}}, \bibinfo {author}
  {\bibfnamefont {Qing}\ \bibnamefont {Zhong}}, \bibinfo {author}
  {\bibfnamefont {Wenbin}\ \bibnamefont {Li}}, \bibinfo {author} {\bibfnamefont
  {Peng}\ \bibnamefont {Cheng}}, \bibinfo {author} {\bibfnamefont {Lan}\
  \bibnamefont {Chen}}, \ and\ \bibinfo {author} {\bibfnamefont {Kehui}\
  \bibnamefont {Wu}},\ }\bibfield  {title} {\enquote {\bibinfo {title}
  {Strain-induced band engineering in monolayer stanene on sb(111)},}\ }\href
  {\doibase 10.1103/physrevmaterials.1.054004} {\bibfield  {journal} {\bibinfo
  {journal} {Physical Review Materials}\ }\textbf {\bibinfo {volume} {1}},\
  \bibinfo {pages} {054004} (\bibinfo {year} {2017})}\BibitemShut {NoStop}%
\bibitem [{\citenamefont {Zhu}\ and\ \citenamefont
  {Tom{\'{a}}nek}(2014)}]{Zhu2014}%
  \BibitemOpen
  \bibfield  {author} {\bibinfo {author} {\bibfnamefont {Zhen}\ \bibnamefont
  {Zhu}}\ and\ \bibinfo {author} {\bibfnamefont {David}\ \bibnamefont
  {Tom{\'{a}}nek}},\ }\bibfield  {title} {\enquote {\bibinfo {title}
  {Semiconducting layered blue phosphorus: A computational study},}\ }\href
  {\doibase 10.1103/physrevlett.112.176802} {\bibfield  {journal} {\bibinfo
  {journal} {Physical Review Letters}\ }\textbf {\bibinfo {volume} {112}},\
  \bibinfo {pages} {176802} (\bibinfo {year} {2014})}\BibitemShut {NoStop}%
\bibitem [{\citenamefont {Zhang}\ \emph {et~al.}(2015)\citenamefont {Zhang},
  \citenamefont {Yan}, \citenamefont {Li}, \citenamefont {Chen},\ and\
  \citenamefont {Zeng}}]{Zhang2015e}%
  \BibitemOpen
  \bibfield  {author} {\bibinfo {author} {\bibfnamefont {Shengli}\ \bibnamefont
  {Zhang}}, \bibinfo {author} {\bibfnamefont {Zhong}\ \bibnamefont {Yan}},
  \bibinfo {author} {\bibfnamefont {Yafei}\ \bibnamefont {Li}}, \bibinfo
  {author} {\bibfnamefont {Zhongfang}\ \bibnamefont {Chen}}, \ and\ \bibinfo
  {author} {\bibfnamefont {Haibo}\ \bibnamefont {Zeng}},\ }\bibfield  {title}
  {\enquote {\bibinfo {title} {Atomically thin arsenene and antimonene:
  Semimetal-semiconductor and indirect-direct band-gap transitions},}\ }\href
  {\doibase 10.1002/anie.201411246} {\bibfield  {journal} {\bibinfo  {journal}
  {Angew. Chem. Int. Ed.}\ }\textbf {\bibinfo {volume} {54}},\ \bibinfo {pages}
  {3112--3115} (\bibinfo {year} {2015})}\BibitemShut {NoStop}%
\bibitem [{\citenamefont {Sharma}\ and\ \citenamefont
  {Pandey}(2018)}]{Sharma2018}%
  \BibitemOpen
  \bibfield  {author} {\bibinfo {author} {\bibfnamefont {Anuj~K.}\ \bibnamefont
  {Sharma}}\ and\ \bibinfo {author} {\bibfnamefont {Ankit~Kumar}\ \bibnamefont
  {Pandey}},\ }\bibfield  {title} {\enquote {\bibinfo {title} {Blue
  phosphorene/{MoS}2heterostructure based {SPR} sensor with enhanced
  sensitivity},}\ }\href {\doibase 10.1109/lpt.2018.2803747} {\bibfield
  {journal} {\bibinfo  {journal} {{IEEE} Photonics Technology Letters}\
  }\textbf {\bibinfo {volume} {30}},\ \bibinfo {pages} {595--598} (\bibinfo
  {year} {2018})}\BibitemShut {NoStop}%
\bibitem [{\citenamefont {Ye}\ \emph {et~al.}(2019)\citenamefont {Ye},
  \citenamefont {Zhu}, \citenamefont {Liu}, \citenamefont {Liu},\ and\
  \citenamefont {Yan}}]{Ye2019}%
  \BibitemOpen
  \bibfield  {author} {\bibinfo {author} {\bibfnamefont {Xiao-Juan}\
  \bibnamefont {Ye}}, \bibinfo {author} {\bibfnamefont {Gui-Lin}\ \bibnamefont
  {Zhu}}, \bibinfo {author} {\bibfnamefont {Jin}\ \bibnamefont {Liu}}, \bibinfo
  {author} {\bibfnamefont {Chun-Sheng}\ \bibnamefont {Liu}}, \ and\ \bibinfo
  {author} {\bibfnamefont {Xiao-Hong}\ \bibnamefont {Yan}},\ }\bibfield
  {title} {\enquote {\bibinfo {title} {Monolayer, bilayer, and heterostructure
  arsenene as potential anode materials for magnesium-ion batteries: A
  first-principles study},}\ }\href {\doibase 10.1021/acs.jpcc.9b02399}
  {\bibfield  {journal} {\bibinfo  {journal} {The Journal of Physical Chemistry
  C}\ }\textbf {\bibinfo {volume} {123}},\ \bibinfo {pages} {15777--15786}
  (\bibinfo {year} {2019})}\BibitemShut {NoStop}%
\bibitem [{\citenamefont {Zhang}\ \emph {et~al.}(2012)\citenamefont {Zhang},
  \citenamefont {Liu}, \citenamefont {Duan}, \citenamefont {Liu},\ and\
  \citenamefont {Wu}}]{Zhang2012}%
  \BibitemOpen
  \bibfield  {author} {\bibinfo {author} {\bibfnamefont {PengFei}\ \bibnamefont
  {Zhang}}, \bibinfo {author} {\bibfnamefont {Zheng}\ \bibnamefont {Liu}},
  \bibinfo {author} {\bibfnamefont {Wenhui}\ \bibnamefont {Duan}}, \bibinfo
  {author} {\bibfnamefont {Feng}\ \bibnamefont {Liu}}, \ and\ \bibinfo {author}
  {\bibfnamefont {Jian}\ \bibnamefont {Wu}},\ }\bibfield  {title} {\enquote
  {\bibinfo {title} {Topological and electronic transitions in a sb(111)
  nanofilm: The interplay between quantum confinement and surface effect},}\
  }\href {\doibase 10.1103/physrevb.85.201410} {\bibfield  {journal} {\bibinfo
  {journal} {Physical Review B}\ }\textbf {\bibinfo {volume} {85}},\ \bibinfo
  {pages} {201410} (\bibinfo {year} {2012})}\BibitemShut {NoStop}%
\bibitem [{\citenamefont {Wang}\ \emph {et~al.}(2017)\citenamefont {Wang},
  \citenamefont {Huang}, \citenamefont {Ye}, \citenamefont {Quhe},
  \citenamefont {Pan}, \citenamefont {Zhang}, \citenamefont {Zhong},
  \citenamefont {Shi},\ and\ \citenamefont {Lu}}]{Wang2017}%
  \BibitemOpen
  \bibfield  {author} {\bibinfo {author} {\bibfnamefont {Yangyang}\
  \bibnamefont {Wang}}, \bibinfo {author} {\bibfnamefont {Pu}~\bibnamefont
  {Huang}}, \bibinfo {author} {\bibfnamefont {Meng}\ \bibnamefont {Ye}},
  \bibinfo {author} {\bibfnamefont {Ruge}\ \bibnamefont {Quhe}}, \bibinfo
  {author} {\bibfnamefont {Yuanyuan}\ \bibnamefont {Pan}}, \bibinfo {author}
  {\bibfnamefont {Han}\ \bibnamefont {Zhang}}, \bibinfo {author} {\bibfnamefont
  {Hongxia}\ \bibnamefont {Zhong}}, \bibinfo {author} {\bibfnamefont {Junjie}\
  \bibnamefont {Shi}}, \ and\ \bibinfo {author} {\bibfnamefont {Jing}\
  \bibnamefont {Lu}},\ }\bibfield  {title} {\enquote {\bibinfo {title}
  {Many-body effect, carrier mobility, and device performance of hexagonal
  arsenene and antimonene},}\ }\href {\doibase 10.1021/acs.chemmater.6b04909}
  {\bibfield  {journal} {\bibinfo  {journal} {Chemistry of Materials}\ }\textbf
  {\bibinfo {volume} {29}},\ \bibinfo {pages} {2191--2201} (\bibinfo {year}
  {2017})}\BibitemShut {NoStop}%
\bibitem [{\citenamefont {Mortazavi}\ \emph {et~al.}(2016)\citenamefont
  {Mortazavi}, \citenamefont {Dianat}, \citenamefont {Cuniberti},\ and\
  \citenamefont {Rabczuk}}]{Mortazavi2016}%
  \BibitemOpen
  \bibfield  {author} {\bibinfo {author} {\bibfnamefont {Bohayra}\ \bibnamefont
  {Mortazavi}}, \bibinfo {author} {\bibfnamefont {Arezoo}\ \bibnamefont
  {Dianat}}, \bibinfo {author} {\bibfnamefont {Gianaurelio}\ \bibnamefont
  {Cuniberti}}, \ and\ \bibinfo {author} {\bibfnamefont {Timon}\ \bibnamefont
  {Rabczuk}},\ }\bibfield  {title} {\enquote {\bibinfo {title} {Application of
  silicene, germanene and stanene for na or li ion storage: A theoretical
  investigation},}\ }\href {\doibase 10.1016/j.electacta.2016.08.027}
  {\bibfield  {journal} {\bibinfo  {journal} {Electrochimica Acta}\ }\textbf
  {\bibinfo {volume} {213}},\ \bibinfo {pages} {865--870} (\bibinfo {year}
  {2016})}\BibitemShut {NoStop}%
\bibitem [{\citenamefont {Kurpas}\ \emph {et~al.}(2019)\citenamefont {Kurpas},
  \citenamefont {Junior}, \citenamefont {Gmitra},\ and\ \citenamefont
  {Fabian}}]{Kurpas2019}%
  \BibitemOpen
  \bibfield  {author} {\bibinfo {author} {\bibfnamefont {Marcin}\ \bibnamefont
  {Kurpas}}, \bibinfo {author} {\bibfnamefont {Paulo E.~Faria}\ \bibnamefont
  {Junior}}, \bibinfo {author} {\bibfnamefont {Martin}\ \bibnamefont {Gmitra}},
  \ and\ \bibinfo {author} {\bibfnamefont {Jaroslav}\ \bibnamefont {Fabian}},\
  }\bibfield  {title} {\enquote {\bibinfo {title} {Spin-orbit coupling in
  elemental two-dimensional materials},}\ }\href {\doibase
  10.1103/physrevb.100.125422} {\bibfield  {journal} {\bibinfo  {journal}
  {Physical Review B}\ }\textbf {\bibinfo {volume} {100}},\ \bibinfo {pages}
  {125422} (\bibinfo {year} {2019})}\BibitemShut {NoStop}%
\bibitem [{\citenamefont {Balendhran}\ \emph {et~al.}(2014)\citenamefont
  {Balendhran}, \citenamefont {Walia}, \citenamefont {Nili}, \citenamefont
  {Sriram},\ and\ \citenamefont {Bhaskaran}}]{Balendhran2014}%
  \BibitemOpen
  \bibfield  {author} {\bibinfo {author} {\bibfnamefont {Sivacarendran}\
  \bibnamefont {Balendhran}}, \bibinfo {author} {\bibfnamefont {Sumeet}\
  \bibnamefont {Walia}}, \bibinfo {author} {\bibfnamefont {Hussein}\
  \bibnamefont {Nili}}, \bibinfo {author} {\bibfnamefont {Sharath}\
  \bibnamefont {Sriram}}, \ and\ \bibinfo {author} {\bibfnamefont {Madhu}\
  \bibnamefont {Bhaskaran}},\ }\bibfield  {title} {\enquote {\bibinfo {title}
  {Elemental analogues of graphene: Silicene, germanene, stanene, and
  phosphorene},}\ }\href {\doibase 10.1002/smll.201402041} {\bibfield
  {journal} {\bibinfo  {journal} {Small}\ }\textbf {\bibinfo {volume} {11}},\
  \bibinfo {pages} {640--652} (\bibinfo {year} {2014})}\BibitemShut {NoStop}%
\bibitem [{\citenamefont {Zhang}\ \emph {et~al.}(2017)\citenamefont {Zhang},
  \citenamefont {Zhang}, \citenamefont {Guo},\ and\ \citenamefont {feng
  Duan}}]{Zhang2017}%
  \BibitemOpen
  \bibfield  {author} {\bibinfo {author} {\bibfnamefont {Dong-Chen}\
  \bibnamefont {Zhang}}, \bibinfo {author} {\bibfnamefont {Ai-Xia}\
  \bibnamefont {Zhang}}, \bibinfo {author} {\bibfnamefont {San-Dong}\
  \bibnamefont {Guo}}, \ and\ \bibinfo {author} {\bibfnamefont
  {Yi}~\bibnamefont {feng Duan}},\ }\bibfield  {title} {\enquote {\bibinfo
  {title} {Thermoelectric properties of beta-as, sb and bi monolayers},}\
  }\href {\doibase 10.1039/c7ra03662k} {\bibfield  {journal} {\bibinfo
  {journal} {{RSC} Advances}\ }\textbf {\bibinfo {volume} {7}},\ \bibinfo
  {pages} {24537--24546} (\bibinfo {year} {2017})}\BibitemShut {NoStop}%
\bibitem [{\citenamefont {Cahangirov}\ \emph {et~al.}(2009)\citenamefont
  {Cahangirov}, \citenamefont {Topsakal}, \citenamefont {Akt眉rk},
  \citenamefont {{\c{S}}ahin},\ and\ \citenamefont {Ciraci}}]{Cahangirov2009}%
  \BibitemOpen
  \bibfield  {author} {\bibinfo {author} {\bibfnamefont {S.}~\bibnamefont
  {Cahangirov}}, \bibinfo {author} {\bibfnamefont {M.}~\bibnamefont
  {Topsakal}}, \bibinfo {author} {\bibfnamefont {E.}~\bibnamefont {Aktuerk}},
  \bibinfo {author} {\bibfnamefont {H.}~\bibnamefont {{\c{S}}ahin}}, \ and\
  \bibinfo {author} {\bibfnamefont {S.}~\bibnamefont {Ciraci}},\ }\bibfield
  {title} {\enquote {\bibinfo {title} {Two- and one-dimensional honeycomb
  structures of silicon and germanium},}\ }\href {\doibase
  10.1103/physrevlett.102.236804} {\bibfield  {journal} {\bibinfo  {journal}
  {Physical Review Letters}\ }\textbf {\bibinfo {volume} {102}},\ \bibinfo
  {pages} {236804} (\bibinfo {year} {2009})}\BibitemShut {NoStop}%
\bibitem [{\citenamefont {Akturk}\ \emph {et~al.}({2016})\citenamefont
  {Akturk}, \citenamefont {Akturk},\ and\ \citenamefont
  {Ciraci}}]{AkturkE2016}%
  \BibitemOpen
  \bibfield  {author} {\bibinfo {author} {\bibfnamefont {E.}~\bibnamefont
  {Akturk}}, \bibinfo {author} {\bibfnamefont {O.~Uzengi}\ \bibnamefont
  {Akturk}}, \ and\ \bibinfo {author} {\bibfnamefont {S.}~\bibnamefont
  {Ciraci}},\ }\bibfield  {title} {\enquote {\bibinfo {title} {{Single and
  bilayer bismuthene: Stability at high temperature and mechanical and
  electronic properties}},}\ }\href {\doibase {10.1103/PhysRevB.94.014115}}
  {\bibfield  {journal} {\bibinfo  {journal} {{PHYSICAL REVIEW B}}\ }\textbf
  {\bibinfo {volume} {{94}}},\ \bibinfo {pages} {014115} (\bibinfo {year}
  {{2016}})}\BibitemShut {NoStop}%
\bibitem [{\citenamefont {SILVI}\ and\ \citenamefont
  {SAVIN}({1994})}]{Silvi1994}%
  \BibitemOpen
  \bibfield  {author} {\bibinfo {author} {\bibfnamefont {B}~\bibnamefont
  {SILVI}}\ and\ \bibinfo {author} {\bibfnamefont {A}~\bibnamefont {SAVIN}},\
  }\bibfield  {title} {\enquote {\bibinfo {title} {{CLASSIFICATION OF
  CHEMICAL-BONDS BASED ON TOPOLOGICAL ANALYSIS OF ELECTRON LOCALIZATION
  FUNCTIONS}},}\ }\href {\doibase {10.1038/371683a0}} {\bibfield  {journal}
  {\bibinfo  {journal} {{NATURE}}\ }\textbf {\bibinfo {volume} {{371}}},\
  \bibinfo {pages} {{683--686}} (\bibinfo {year} {{1994}})}\BibitemShut
  {NoStop}%
\bibitem [{\citenamefont {Jahn}\ and\ \citenamefont {Teller}(1937)}]{1937a}%
  \BibitemOpen
  \bibfield  {author} {\bibinfo {author} {\bibfnamefont {H.~A.}\ \bibnamefont
  {Jahn}}\ and\ \bibinfo {author} {\bibfnamefont {E.}~\bibnamefont {Teller}},\
  }\bibfield  {title} {\enquote {\bibinfo {title} {Stability of polyatomic
  molecules in degenerate electronic states - i{\textemdash}orbital
  degeneracy},}\ }\href {\doibase 10.1098/rspa.1937.0142} {\bibfield  {journal}
  {\bibinfo  {journal} {Proceedings of the Royal Society of London. Series A -
  Mathematical and Physical Sciences}\ }\textbf {\bibinfo {volume} {161}},\
  \bibinfo {pages} {220--235} (\bibinfo {year} {1937})}\BibitemShut {NoStop}%
\bibitem [{\citenamefont {Shao}\ \emph {et~al.}(2016)\citenamefont {Shao},
  \citenamefont {Tan}, \citenamefont {Jiang},\ and\ \citenamefont
  {Jiang}}]{Shao2016}%
  \BibitemOpen
  \bibfield  {author} {\bibinfo {author} {\bibfnamefont {Hezhu}\ \bibnamefont
  {Shao}}, \bibinfo {author} {\bibfnamefont {Xiaojian}\ \bibnamefont {Tan}},
  \bibinfo {author} {\bibfnamefont {Jun}\ \bibnamefont {Jiang}}, \ and\
  \bibinfo {author} {\bibfnamefont {Haochuan}\ \bibnamefont {Jiang}},\
  }\bibfield  {title} {\enquote {\bibinfo {title} {First-principles study on
  the elastic properties of cu2gese3},}\ }\href {\doibase
  10.1209/0295-5075/113/26001} {\bibfield  {journal} {\bibinfo  {journal}
  {{EPL} (Europhysics Letters)}\ }\textbf {\bibinfo {volume} {113}},\ \bibinfo
  {pages} {26001} (\bibinfo {year} {2016})}\BibitemShut {NoStop}%
\bibitem [{\citenamefont {Lang}\ \emph {et~al.}(2016)\citenamefont {Lang},
  \citenamefont {Zhang},\ and\ \citenamefont {Liu}}]{Lang2016}%
  \BibitemOpen
  \bibfield  {author} {\bibinfo {author} {\bibfnamefont {Haifeng}\ \bibnamefont
  {Lang}}, \bibinfo {author} {\bibfnamefont {Shuqing}\ \bibnamefont {Zhang}}, \
  and\ \bibinfo {author} {\bibfnamefont {Zhirong}\ \bibnamefont {Liu}},\
  }\bibfield  {title} {\enquote {\bibinfo {title} {Mobility anisotropy of
  two-dimensional semiconductors},}\ }\href {\doibase
  10.1103/physrevb.94.235306} {\bibfield  {journal} {\bibinfo  {journal}
  {Physical Review B}\ }\textbf {\bibinfo {volume} {94}},\ \bibinfo {pages}
  {235306} (\bibinfo {year} {2016})}\BibitemShut {NoStop}%
\bibitem [{\citenamefont {Herring}\ and\ \citenamefont
  {Vogt}(1956)}]{Herring1956}%
  \BibitemOpen
  \bibfield  {author} {\bibinfo {author} {\bibfnamefont {Conyers}\ \bibnamefont
  {Herring}}\ and\ \bibinfo {author} {\bibfnamefont {Erich}\ \bibnamefont
  {Vogt}},\ }\bibfield  {title} {\enquote {\bibinfo {title} {Transport and
  deformation-potential theory for many-valley semiconductors with anisotropic
  scattering},}\ }\href {\doibase 10.1103/PhysRev.101.944} {\bibfield
  {journal} {\bibinfo  {journal} {Phys. Rev.}\ }\textbf {\bibinfo {volume}
  {101}},\ \bibinfo {pages} {944--961} (\bibinfo {year} {1956})}\BibitemShut
  {NoStop}%
\bibitem [{\citenamefont {Malard}\ \emph {et~al.}(2009)\citenamefont {Malard},
  \citenamefont {Guimar\~aes}, \citenamefont {Mafra}, \citenamefont {Mazzoni},\
  and\ \citenamefont {Jorio}}]{Marlard2009}%
  \BibitemOpen
  \bibfield  {author} {\bibinfo {author} {\bibfnamefont {L.~M.}\ \bibnamefont
  {Malard}}, \bibinfo {author} {\bibfnamefont {M.~H.~D.}\ \bibnamefont
  {Guimar\~aes}}, \bibinfo {author} {\bibfnamefont {D.~L.}\ \bibnamefont
  {Mafra}}, \bibinfo {author} {\bibfnamefont {M.~S.~C.}\ \bibnamefont
  {Mazzoni}}, \ and\ \bibinfo {author} {\bibfnamefont {A.}~\bibnamefont
  {Jorio}},\ }\bibfield  {title} {\enquote {\bibinfo {title} {Group-theory
  analysis of electrons and phonons in $n$-layer graphene systems},}\ }\href
  {\doibase 10.1103/PhysRevB.79.125426} {\bibfield  {journal} {\bibinfo
  {journal} {Phys. Rev. B}\ }\textbf {\bibinfo {volume} {79}},\ \bibinfo
  {pages} {125426} (\bibinfo {year} {2009})}\BibitemShut {NoStop}%
\bibitem [{\citenamefont {Jiang}\ \emph {et~al.}(2005)\citenamefont {Jiang},
  \citenamefont {Saito}, \citenamefont {Gr\"uneis}, \citenamefont {Chou},
  \citenamefont {Samsonidze}, \citenamefont {Jorio}, \citenamefont
  {Dresselhaus},\ and\ \citenamefont {Dresselhaus}}]{jiang2005}%
  \BibitemOpen
  \bibfield  {author} {\bibinfo {author} {\bibfnamefont {J.}~\bibnamefont
  {Jiang}}, \bibinfo {author} {\bibfnamefont {R.}~\bibnamefont {Saito}},
  \bibinfo {author} {\bibfnamefont {A.}~\bibnamefont {Gr\"uneis}}, \bibinfo
  {author} {\bibfnamefont {S.~G.}\ \bibnamefont {Chou}}, \bibinfo {author}
  {\bibfnamefont {Ge.~G.}\ \bibnamefont {Samsonidze}}, \bibinfo {author}
  {\bibfnamefont {A.}~\bibnamefont {Jorio}}, \bibinfo {author} {\bibfnamefont
  {G.}~\bibnamefont {Dresselhaus}}, \ and\ \bibinfo {author} {\bibfnamefont
  {M.~S.}\ \bibnamefont {Dresselhaus}},\ }\bibfield  {title} {\enquote
  {\bibinfo {title} {Intensity of the resonance raman excitation spectra of
  single-wall carbon nanotubes},}\ }\href {\doibase 10.1103/PhysRevB.71.205420}
  {\bibfield  {journal} {\bibinfo  {journal} {Phys. Rev. B}\ }\textbf {\bibinfo
  {volume} {71}},\ \bibinfo {pages} {205420} (\bibinfo {year}
  {2005})}\BibitemShut {NoStop}%
\bibitem [{\citenamefont {Castro~Neto}\ and\ \citenamefont
  {Guinea}(2007)}]{Castro2007}%
  \BibitemOpen
  \bibfield  {author} {\bibinfo {author} {\bibfnamefont {A.~H.}\ \bibnamefont
  {Castro~Neto}}\ and\ \bibinfo {author} {\bibfnamefont {Francisco}\
  \bibnamefont {Guinea}},\ }\bibfield  {title} {\enquote {\bibinfo {title}
  {Electron-phonon coupling and raman spectroscopy in graphene},}\ }\href
  {\doibase 10.1103/PhysRevB.75.045404} {\bibfield  {journal} {\bibinfo
  {journal} {Phys. Rev. B}\ }\textbf {\bibinfo {volume} {75}},\ \bibinfo
  {pages} {045404} (\bibinfo {year} {2007})}\BibitemShut {NoStop}%
\bibitem [{\citenamefont {Chu}\ \emph {et~al.}(2014)\citenamefont {Chu},
  \citenamefont {Gautreau},\ and\ \citenamefont {Basaran}}]{Chu2014Parity}%
  \BibitemOpen
  \bibfield  {author} {\bibinfo {author} {\bibfnamefont {Yanbiao}\ \bibnamefont
  {Chu}}, \bibinfo {author} {\bibfnamefont {Pierre}\ \bibnamefont {Gautreau}},
  \ and\ \bibinfo {author} {\bibfnamefont {Cemal}\ \bibnamefont {Basaran}},\
  }\bibfield  {title} {\enquote {\bibinfo {title} {Parity conservation in
  electron-phonon scattering in zigzag graphene nanoribbon},}\ }\href@noop {}
  {\bibfield  {journal} {\bibinfo  {journal} {Applied Physics Letters}\
  }\textbf {\bibinfo {volume} {105}},\ \bibinfo {pages} {347} (\bibinfo {year}
  {2014})}\BibitemShut {NoStop}%
\bibitem [{\citenamefont {Dresselhaus}\ \emph {et~al.}(2008)\citenamefont
  {Dresselhaus}, \citenamefont {Dresselhaus},\ and\ \citenamefont
  {Jorio}}]{Dresselhaus2008}%
  \BibitemOpen
  \bibfield  {author} {\bibinfo {author} {\bibfnamefont {Mildred~S.}\
  \bibnamefont {Dresselhaus}}, \bibinfo {author} {\bibfnamefont {Gene}\
  \bibnamefont {Dresselhaus}}, \ and\ \bibinfo {author} {\bibfnamefont {Ado}\
  \bibnamefont {Jorio}},\ }\href@noop {} {\emph {\bibinfo {title} {Group
  Theory}}}\ (\bibinfo  {publisher} {Springer-Verlag New York, LLC},\ \bibinfo
  {year} {2008})\BibitemShut {NoStop}%
\bibitem [{\citenamefont {Pei}\ \emph {et~al.}(2011)\citenamefont {Pei},
  \citenamefont {Shi}, \citenamefont {LaLonde}, \citenamefont {Wang},
  \citenamefont {Chen},\ and\ \citenamefont {Snyder}}]{Pei2011}%
  \BibitemOpen
  \bibfield  {author} {\bibinfo {author} {\bibfnamefont {Yanzhong}\
  \bibnamefont {Pei}}, \bibinfo {author} {\bibfnamefont {Xiaoya}\ \bibnamefont
  {Shi}}, \bibinfo {author} {\bibfnamefont {Aaron}\ \bibnamefont {LaLonde}},
  \bibinfo {author} {\bibfnamefont {Heng}\ \bibnamefont {Wang}}, \bibinfo
  {author} {\bibfnamefont {Lidong}\ \bibnamefont {Chen}}, \ and\ \bibinfo
  {author} {\bibfnamefont {G.~Jeffrey}\ \bibnamefont {Snyder}},\ }\bibfield
  {title} {\enquote {\bibinfo {title} {Convergence of electronic bands for high
  performance bulk thermoelectrics},}\ }\href {\doibase 10.1038/nature09996}
  {\bibfield  {journal} {\bibinfo  {journal} {Nature}\ }\textbf {\bibinfo
  {volume} {473}},\ \bibinfo {pages} {66--69} (\bibinfo {year}
  {2011})}\BibitemShut {NoStop}%
\bibitem [{\citenamefont {Zhao}\ \emph {et~al.}(2016)\citenamefont {Zhao},
  \citenamefont {Chang}, \citenamefont {Tan},\ and\ \citenamefont
  {Kanatzidis}}]{lidong2016}%
  \BibitemOpen
  \bibfield  {author} {\bibinfo {author} {\bibfnamefont {Li-Dong}\ \bibnamefont
  {Zhao}}, \bibinfo {author} {\bibfnamefont {Cheng}\ \bibnamefont {Chang}},
  \bibinfo {author} {\bibfnamefont {Gangjian}\ \bibnamefont {Tan}}, \ and\
  \bibinfo {author} {\bibfnamefont {Mercouri~G.}\ \bibnamefont {Kanatzidis}},\
  }\bibfield  {title} {\enquote {\bibinfo {title} {Snse: a remarkable new
  thermoelectric material},}\ }\href {\doibase 10.1039/C6EE01755J} {\bibfield
  {journal} {\bibinfo  {journal} {Energy Environ. Sci.}\ }\textbf {\bibinfo
  {volume} {9}},\ \bibinfo {pages} {3044--3060} (\bibinfo {year}
  {2016})}\BibitemShut {NoStop}%
\bibitem [{\citenamefont {Chen}\ \emph {et~al.}(2017)\citenamefont {Chen},
  \citenamefont {Lyu}, \citenamefont {Wang}, \citenamefont {Fu}, \citenamefont
  {Heng},\ and\ \citenamefont {Mo}}]{Chen2017}%
  \BibitemOpen
  \bibfield  {author} {\bibinfo {author} {\bibfnamefont {Kai-Xuan}\
  \bibnamefont {Chen}}, \bibinfo {author} {\bibfnamefont {Shu-Shen}\
  \bibnamefont {Lyu}}, \bibinfo {author} {\bibfnamefont {Xiao-Ming}\
  \bibnamefont {Wang}}, \bibinfo {author} {\bibfnamefont {Yuan-Xiang}\
  \bibnamefont {Fu}}, \bibinfo {author} {\bibfnamefont {Yi}~\bibnamefont
  {Heng}}, \ and\ \bibinfo {author} {\bibfnamefont {Dong-Chuan}\ \bibnamefont
  {Mo}},\ }\bibfield  {title} {\enquote {\bibinfo {title} {Excellent
  thermoelectric performance predicted in two-dimensional buckled antimonene: A
  first-principles study},}\ }\href {\doibase 10.1021/acs.jpcc.7b03129}
  {\bibfield  {journal} {\bibinfo  {journal} {The Journal of Physical Chemistry
  C}\ }\textbf {\bibinfo {volume} {121}},\ \bibinfo {pages} {13035--13042}
  (\bibinfo {year} {2017})}\BibitemShut {NoStop}%
\bibitem [{\citenamefont {Sun}\ \emph {et~al.}(2015)\citenamefont {Sun},
  \citenamefont {Wei}, \citenamefont {Zhang}, \citenamefont {Tomczak},
  \citenamefont {Strydom}, \citenamefont {S{\o}ndergaard}, \citenamefont
  {Iversen},\ and\ \citenamefont {Steglich}}]{Sun2015}%
  \BibitemOpen
  \bibfield  {author} {\bibinfo {author} {\bibfnamefont {Peijie}\ \bibnamefont
  {Sun}}, \bibinfo {author} {\bibfnamefont {Beipei}\ \bibnamefont {Wei}},
  \bibinfo {author} {\bibfnamefont {Jiahao}\ \bibnamefont {Zhang}}, \bibinfo
  {author} {\bibfnamefont {Jan~M.}\ \bibnamefont {Tomczak}}, \bibinfo {author}
  {\bibfnamefont {A.M.}\ \bibnamefont {Strydom}}, \bibinfo {author}
  {\bibfnamefont {M.}~\bibnamefont {S{\o}ndergaard}}, \bibinfo {author}
  {\bibfnamefont {Bo~B.}\ \bibnamefont {Iversen}}, \ and\ \bibinfo {author}
  {\bibfnamefont {Frank}\ \bibnamefont {Steglich}},\ }\bibfield  {title}
  {\enquote {\bibinfo {title} {Large seebeck effect by charge-mobility
  engineering},}\ }\href {\doibase 10.1038/ncomms8475} {\bibfield  {journal}
  {\bibinfo  {journal} {Nature Communications}\ }\textbf {\bibinfo {volume}
  {6}} (\bibinfo {year} {2015}),\ 10.1038/ncomms8475}\BibitemShut {NoStop}%
\bibitem [{\citenamefont {Wei}\ \emph {et~al.}(2015)\citenamefont {Wei},
  \citenamefont {Zhang}, \citenamefont {Sun}, \citenamefont {Wang},
  \citenamefont {Wang},\ and\ \citenamefont {Steglich}}]{Wei2015}%
  \BibitemOpen
  \bibfield  {author} {\bibinfo {author} {\bibfnamefont {Beipei}\ \bibnamefont
  {Wei}}, \bibinfo {author} {\bibfnamefont {Jiahao}\ \bibnamefont {Zhang}},
  \bibinfo {author} {\bibfnamefont {Peijie}\ \bibnamefont {Sun}}, \bibinfo
  {author} {\bibfnamefont {Wenquan}\ \bibnamefont {Wang}}, \bibinfo {author}
  {\bibfnamefont {Nanlin}\ \bibnamefont {Wang}}, \ and\ \bibinfo {author}
  {\bibfnamefont {Frank}\ \bibnamefont {Steglich}},\ }\bibfield  {title}
  {\enquote {\bibinfo {title} {Nernst effect of the intermediate valence
  compound {YbAl}3: revisiting the thermoelectric properties},}\ }\href
  {\doibase 10.1088/0953-8984/27/10/105601} {\bibfield  {journal} {\bibinfo
  {journal} {Journal of Physics: Condensed Matter}\ }\textbf {\bibinfo {volume}
  {27}},\ \bibinfo {pages} {105601} (\bibinfo {year} {2015})}\BibitemShut
  {NoStop}%
\bibitem [{\citenamefont {Liang}\ \emph {et~al.}()\citenamefont {Liang},
  \citenamefont {Fan}, \citenamefont {Jiang}, \citenamefont {Liu},\ and\
  \citenamefont {Zhao}}]{Liang2016}%
  \BibitemOpen
  \bibfield  {author} {\bibinfo {author} {\bibfnamefont {Jinghua}\ \bibnamefont
  {Liang}}, \bibinfo {author} {\bibfnamefont {Dengdong}\ \bibnamefont {Fan}},
  \bibinfo {author} {\bibfnamefont {Peiheng}\ \bibnamefont {Jiang}}, \bibinfo
  {author} {\bibfnamefont {Huijun}\ \bibnamefont {Liu}}, \ and\ \bibinfo
  {author} {\bibfnamefont {Wenyu}\ \bibnamefont {Zhao}},\ }\bibfield  {title}
  {\enquote {\bibinfo {title} {Phonon-limited electrical transport properties
  of intermetallic compound ybal3 from first-principles calculations},}\
  }\href@noop {} {\ }\Eprint
  {http://arxiv.org/abs/http://arxiv.org/abs/1609.05858v1}
  {http://arxiv.org/abs/1609.05858v1} \BibitemShut {NoStop}%
\bibitem [{\citenamefont {Poudel}\ \emph {et~al.}({2008})\citenamefont
  {Poudel}, \citenamefont {Hao}, \citenamefont {Ma}, \citenamefont {Lan},
  \citenamefont {Minnich}, \citenamefont {Yu}, \citenamefont {Yan},
  \citenamefont {Wang}, \citenamefont {Muto}, \citenamefont {Vashaee},
  \citenamefont {Chen}, \citenamefont {Liu}, \citenamefont {Dresselhaus},
  \citenamefont {Chen},\ and\ \citenamefont {Ren}}]{Poudel2008}%
  \BibitemOpen
  \bibfield  {author} {\bibinfo {author} {\bibfnamefont {Bed}\ \bibnamefont
  {Poudel}}, \bibinfo {author} {\bibfnamefont {Qing}\ \bibnamefont {Hao}},
  \bibinfo {author} {\bibfnamefont {Yi}~\bibnamefont {Ma}}, \bibinfo {author}
  {\bibfnamefont {Yucheng}\ \bibnamefont {Lan}}, \bibinfo {author}
  {\bibfnamefont {Austin}\ \bibnamefont {Minnich}}, \bibinfo {author}
  {\bibfnamefont {Bo}~\bibnamefont {Yu}}, \bibinfo {author} {\bibfnamefont
  {Xiao}\ \bibnamefont {Yan}}, \bibinfo {author} {\bibfnamefont {Dezhi}\
  \bibnamefont {Wang}}, \bibinfo {author} {\bibfnamefont {Andrew}\ \bibnamefont
  {Muto}}, \bibinfo {author} {\bibfnamefont {Daryoosh}\ \bibnamefont
  {Vashaee}}, \bibinfo {author} {\bibfnamefont {Xiaoyuan}\ \bibnamefont
  {Chen}}, \bibinfo {author} {\bibfnamefont {Junming}\ \bibnamefont {Liu}},
  \bibinfo {author} {\bibfnamefont {Mildred~S.}\ \bibnamefont {Dresselhaus}},
  \bibinfo {author} {\bibfnamefont {Gang}\ \bibnamefont {Chen}}, \ and\
  \bibinfo {author} {\bibfnamefont {Zhifeng}\ \bibnamefont {Ren}},\ }\bibfield
  {title} {\enquote {\bibinfo {title} {{High-thermoelectric performance of
  nanostructured bismuth antimony telluride bulk alloys}},}\ }\href {\doibase
  {10.1126/science.1156446}} {\bibfield  {journal} {\bibinfo  {journal}
  {{SCIENCE}}\ }\textbf {\bibinfo {volume} {{320}}},\ \bibinfo {pages}
  {{634--638}} (\bibinfo {year} {{2008}})}\BibitemShut {NoStop}%
\bibitem [{\citenamefont {Nielsen}\ \emph {et~al.}({2012})\citenamefont
  {Nielsen}, \citenamefont {Levin}, \citenamefont {Jaworski}, \citenamefont
  {Schmidt-Rohr},\ and\ \citenamefont {Heremans}}]{Nielsen2012}%
  \BibitemOpen
  \bibfield  {author} {\bibinfo {author} {\bibfnamefont {M.~D.}\ \bibnamefont
  {Nielsen}}, \bibinfo {author} {\bibfnamefont {E.~M.}\ \bibnamefont {Levin}},
  \bibinfo {author} {\bibfnamefont {C.~M.}\ \bibnamefont {Jaworski}}, \bibinfo
  {author} {\bibfnamefont {K.}~\bibnamefont {Schmidt-Rohr}}, \ and\ \bibinfo
  {author} {\bibfnamefont {J.~P.}\ \bibnamefont {Heremans}},\ }\bibfield
  {title} {\enquote {\bibinfo {title} {{Chromium as resonant donor impurity in
  PbTe}},}\ }\href {\doibase {10.1103/PhysRevB.85.045210}} {\bibfield
  {journal} {\bibinfo  {journal} {{PHYSICAL REVIEW B}}\ }\textbf {\bibinfo
  {volume} {{85}}},\ \bibinfo {pages} {045210} (\bibinfo {year}
  {{2012}})}\BibitemShut {NoStop}%
\bibitem [{\citenamefont {Zhang}\ and\ \citenamefont {Zhao}(2015)}]{zhao2015}%
  \BibitemOpen
  \bibfield  {author} {\bibinfo {author} {\bibfnamefont {Xiao}\ \bibnamefont
  {Zhang}}\ and\ \bibinfo {author} {\bibfnamefont {Li-Dong}\ \bibnamefont
  {Zhao}},\ }\bibfield  {title} {\enquote {\bibinfo {title} {Thermoelectric
  materials: Energy conversion between heat and electricity},}\ }\href
  {\doibase https://doi.org/10.1016/j.jmat.2015.01.001} {\bibfield  {journal}
  {\bibinfo  {journal} {Journal of Materiomics}\ }\textbf {\bibinfo {volume}
  {1}},\ \bibinfo {pages} {92 -- 105} (\bibinfo {year} {2015})}\BibitemShut
  {NoStop}%
\bibitem [{\citenamefont {Peng}\ \emph {et~al.}(2017)\citenamefont {Peng},
  \citenamefont {Zhang}, \citenamefont {Zhang}, \citenamefont {Shao},
  \citenamefont {Ni}, \citenamefont {Zhu},\ and\ \citenamefont
  {Zhu}}]{Peng2017}%
  \BibitemOpen
  \bibfield  {author} {\bibinfo {author} {\bibfnamefont {Bo}~\bibnamefont
  {Peng}}, \bibinfo {author} {\bibfnamefont {Dequan}\ \bibnamefont {Zhang}},
  \bibinfo {author} {\bibfnamefont {Hao}\ \bibnamefont {Zhang}}, \bibinfo
  {author} {\bibfnamefont {Hezhu}\ \bibnamefont {Shao}}, \bibinfo {author}
  {\bibfnamefont {Gang}\ \bibnamefont {Ni}}, \bibinfo {author} {\bibfnamefont
  {Yongyuan}\ \bibnamefont {Zhu}}, \ and\ \bibinfo {author} {\bibfnamefont
  {Heyuan}\ \bibnamefont {Zhu}},\ }\bibfield  {title} {\enquote {\bibinfo
  {title} {The conflicting role of buckled structure in phonon transport of 2d
  group-iv and group-v materials},}\ }\href {\doibase 10.1039/C7NR00838D}
  {\bibfield  {journal} {\bibinfo  {journal} {Nanoscale}\ }\textbf {\bibinfo
  {volume} {9}},\ \bibinfo {pages} {7397} (\bibinfo {year} {2017})}\BibitemShut
  {NoStop}%
\bibitem [{\citenamefont {Zhang}\ \emph {et~al.}({2013})\citenamefont {Zhang},
  \citenamefont {Yang}, \citenamefont {Zhang}, \citenamefont {Chen},
  \citenamefont {Liu}, \citenamefont {Wang}, \citenamefont {Tian},
  \citenamefont {Broido}, \citenamefont {Chen},\ and\ \citenamefont
  {Ren}}]{Qinyong2013}%
  \BibitemOpen
  \bibfield  {author} {\bibinfo {author} {\bibfnamefont {Qinyong}\ \bibnamefont
  {Zhang}}, \bibinfo {author} {\bibfnamefont {Siqi}\ \bibnamefont {Yang}},
  \bibinfo {author} {\bibfnamefont {Qian}\ \bibnamefont {Zhang}}, \bibinfo
  {author} {\bibfnamefont {Shuo}\ \bibnamefont {Chen}}, \bibinfo {author}
  {\bibfnamefont {Weishu}\ \bibnamefont {Liu}}, \bibinfo {author}
  {\bibfnamefont {Hui}\ \bibnamefont {Wang}}, \bibinfo {author} {\bibfnamefont
  {Zhiting}\ \bibnamefont {Tian}}, \bibinfo {author} {\bibfnamefont {David}\
  \bibnamefont {Broido}}, \bibinfo {author} {\bibfnamefont {Gang}\ \bibnamefont
  {Chen}}, \ and\ \bibinfo {author} {\bibfnamefont {Zhifeng}\ \bibnamefont
  {Ren}},\ }\bibfield  {title} {\enquote {\bibinfo {title} {{Effect of aluminum
  on the thermoelectric properties of nanostructured PbTe}},}\ }\href {\doibase
  {10.1088/0957-4484/24/34/345705}} {\bibfield  {journal} {\bibinfo  {journal}
  {{NANOTECHNOLOGY}}\ }\textbf {\bibinfo {volume} {{24}}},\ \bibinfo {pages}
  {345705} (\bibinfo {year} {{2013}})}\BibitemShut {NoStop}%
\bibitem [{\citenamefont {Wang}\ \emph {et~al.}(2015)\citenamefont {Wang},
  \citenamefont {Zhang}, \citenamefont {Yu},\ and\ \citenamefont
  {Wang}}]{Wang2015d}%
  \BibitemOpen
  \bibfield  {author} {\bibinfo {author} {\bibfnamefont {Fancy~Qian}\
  \bibnamefont {Wang}}, \bibinfo {author} {\bibfnamefont {Shunhong}\
  \bibnamefont {Zhang}}, \bibinfo {author} {\bibfnamefont {Jiabing}\
  \bibnamefont {Yu}}, \ and\ \bibinfo {author} {\bibfnamefont {Qian}\
  \bibnamefont {Wang}},\ }\bibfield  {title} {\enquote {\bibinfo {title}
  {Thermoelectric properties of single-layered snse sheet},}\ }\href {\doibase
  10.1039/C5NR03813H} {\bibfield  {journal} {\bibinfo  {journal} {Nanoscale}\
  }\textbf {\bibinfo {volume} {7}},\ \bibinfo {pages} {15962--15970} (\bibinfo
  {year} {2015})}\BibitemShut {NoStop}%
\bibitem [{\citenamefont {Wang}\ \emph {et~al.}(2014)\citenamefont {Wang},
  \citenamefont {Gibbs}, \citenamefont {Takagiwa},\ and\ \citenamefont
  {Snyder}}]{Wang2014}%
  \BibitemOpen
  \bibfield  {author} {\bibinfo {author} {\bibfnamefont {Heng}\ \bibnamefont
  {Wang}}, \bibinfo {author} {\bibfnamefont {Zachary~M.}\ \bibnamefont
  {Gibbs}}, \bibinfo {author} {\bibfnamefont {Yoshiki}\ \bibnamefont
  {Takagiwa}}, \ and\ \bibinfo {author} {\bibfnamefont {G.~Jeffrey}\
  \bibnamefont {Snyder}},\ }\bibfield  {title} {\enquote {\bibinfo {title}
  {Tuning bands of {PbSe} for better thermoelectric efficiency},}\ }\href
  {\doibase 10.1039/c3ee43438a} {\bibfield  {journal} {\bibinfo  {journal}
  {Energy Environ. Sci.}\ }\textbf {\bibinfo {volume} {7}},\ \bibinfo {pages}
  {804--811} (\bibinfo {year} {2014})}\BibitemShut {NoStop}%
\bibitem [{\citenamefont {Liu}\ \emph {et~al.}(2012)\citenamefont {Liu},
  \citenamefont {Shi}, \citenamefont {Xu}, \citenamefont {Zhang}, \citenamefont
  {Zhang}, \citenamefont {Chen}, \citenamefont {Li}, \citenamefont {Uher},
  \citenamefont {Day},\ and\ \citenamefont {Snyder}}]{Liu2012}%
  \BibitemOpen
  \bibfield  {author} {\bibinfo {author} {\bibfnamefont {Huili}\ \bibnamefont
  {Liu}}, \bibinfo {author} {\bibfnamefont {Xun}\ \bibnamefont {Shi}}, \bibinfo
  {author} {\bibfnamefont {Fangfang}\ \bibnamefont {Xu}}, \bibinfo {author}
  {\bibfnamefont {Linlin}\ \bibnamefont {Zhang}}, \bibinfo {author}
  {\bibfnamefont {Wenqing}\ \bibnamefont {Zhang}}, \bibinfo {author}
  {\bibfnamefont {Lidong}\ \bibnamefont {Chen}}, \bibinfo {author}
  {\bibfnamefont {Qiang}\ \bibnamefont {Li}}, \bibinfo {author} {\bibfnamefont
  {Ctirad}\ \bibnamefont {Uher}}, \bibinfo {author} {\bibfnamefont {Tristan}\
  \bibnamefont {Day}}, \ and\ \bibinfo {author} {\bibfnamefont {G.~Jeffrey}\
  \bibnamefont {Snyder}},\ }\bibfield  {title} {\enquote {\bibinfo {title}
  {Copper ion liquid-like thermoelectrics},}\ }\href {\doibase
  10.1038/nmat3273} {\bibfield  {journal} {\bibinfo  {journal} {Nature
  Materials}\ }\textbf {\bibinfo {volume} {11}},\ \bibinfo {pages} {422--425}
  (\bibinfo {year} {2012})}\BibitemShut {NoStop}%
\bibitem [{\citenamefont {Ahn}\ \emph {et~al.}(2013)\citenamefont {Ahn},
  \citenamefont {Biswas}, \citenamefont {He}, \citenamefont {Chung},
  \citenamefont {Dravid},\ and\ \citenamefont {Kanatzidis}}]{Ahn2013}%
  \BibitemOpen
  \bibfield  {author} {\bibinfo {author} {\bibfnamefont {Kyunghan}\
  \bibnamefont {Ahn}}, \bibinfo {author} {\bibfnamefont {Kanishka}\
  \bibnamefont {Biswas}}, \bibinfo {author} {\bibfnamefont {Jiaqing}\
  \bibnamefont {He}}, \bibinfo {author} {\bibfnamefont {In}~\bibnamefont
  {Chung}}, \bibinfo {author} {\bibfnamefont {Vinayak}\ \bibnamefont {Dravid}},
  \ and\ \bibinfo {author} {\bibfnamefont {Mercouri~G.}\ \bibnamefont
  {Kanatzidis}},\ }\bibfield  {title} {\enquote {\bibinfo {title} {Enhanced
  thermoelectric properties of p-type nanostructured {PbTe}{\textendash}{MTe}
  (m = cd, hg) materials},}\ }\href {\doibase 10.1039/c3ee40482j} {\bibfield
  {journal} {\bibinfo  {journal} {Energy {\&} Environmental Science}\ }\textbf
  {\bibinfo {volume} {6}},\ \bibinfo {pages} {1529} (\bibinfo {year}
  {2013})}\BibitemShut {NoStop}%
\bibitem [{\citenamefont {Giannozzi}\ \emph {et~al.}(2009)\citenamefont
  {Giannozzi}, \citenamefont {Baroni}, \citenamefont {Bonini}, \citenamefont
  {Calandra}, \citenamefont {Car}, \citenamefont {Cavazzoni}, \citenamefont
  {Ceresoli}, \citenamefont {Chiarotti}, \citenamefont {Cococcioni},
  \citenamefont {Dabo}, \citenamefont {Corso}, \citenamefont {de~Gironcoli},
  \citenamefont {Fabris}, \citenamefont {Fratesi}, \citenamefont {Gebauer},
  \citenamefont {Gerstmann}, \citenamefont {Gougoussis}, \citenamefont
  {Kokalj}, \citenamefont {Lazzeri}, \citenamefont {Martin-Samos},
  \citenamefont {Marzari}, \citenamefont {Mauri}, \citenamefont {Mazzarello},
  \citenamefont {Paolini}, \citenamefont {Pasquarello}, \citenamefont
  {Paulatto}, \citenamefont {Sbraccia}, \citenamefont {Scandolo}, \citenamefont
  {Sclauzero}, \citenamefont {Seitsonen}, \citenamefont {Smogunov},
  \citenamefont {Umari},\ and\ \citenamefont {Wentzcovitch}}]{Giannozzi2009}%
  \BibitemOpen
  \bibfield  {author} {\bibinfo {author} {\bibfnamefont {Paolo}\ \bibnamefont
  {Giannozzi}}, \bibinfo {author} {\bibfnamefont {Stefano}\ \bibnamefont
  {Baroni}}, \bibinfo {author} {\bibfnamefont {Nicola}\ \bibnamefont {Bonini}},
  \bibinfo {author} {\bibfnamefont {Matteo}\ \bibnamefont {Calandra}}, \bibinfo
  {author} {\bibfnamefont {Roberto}\ \bibnamefont {Car}}, \bibinfo {author}
  {\bibfnamefont {Carlo}\ \bibnamefont {Cavazzoni}}, \bibinfo {author}
  {\bibfnamefont {Davide}\ \bibnamefont {Ceresoli}}, \bibinfo {author}
  {\bibfnamefont {Guido~L}\ \bibnamefont {Chiarotti}}, \bibinfo {author}
  {\bibfnamefont {Matteo}\ \bibnamefont {Cococcioni}}, \bibinfo {author}
  {\bibfnamefont {Ismaila}\ \bibnamefont {Dabo}}, \bibinfo {author}
  {\bibfnamefont {Andrea~Dal}\ \bibnamefont {Corso}}, \bibinfo {author}
  {\bibfnamefont {Stefano}\ \bibnamefont {de~Gironcoli}}, \bibinfo {author}
  {\bibfnamefont {Stefano}\ \bibnamefont {Fabris}}, \bibinfo {author}
  {\bibfnamefont {Guido}\ \bibnamefont {Fratesi}}, \bibinfo {author}
  {\bibfnamefont {Ralph}\ \bibnamefont {Gebauer}}, \bibinfo {author}
  {\bibfnamefont {Uwe}\ \bibnamefont {Gerstmann}}, \bibinfo {author}
  {\bibfnamefont {Christos}\ \bibnamefont {Gougoussis}}, \bibinfo {author}
  {\bibfnamefont {Anton}\ \bibnamefont {Kokalj}}, \bibinfo {author}
  {\bibfnamefont {Michele}\ \bibnamefont {Lazzeri}}, \bibinfo {author}
  {\bibfnamefont {Layla}\ \bibnamefont {Martin-Samos}}, \bibinfo {author}
  {\bibfnamefont {Nicola}\ \bibnamefont {Marzari}}, \bibinfo {author}
  {\bibfnamefont {Francesco}\ \bibnamefont {Mauri}}, \bibinfo {author}
  {\bibfnamefont {Riccardo}\ \bibnamefont {Mazzarello}}, \bibinfo {author}
  {\bibfnamefont {Stefano}\ \bibnamefont {Paolini}}, \bibinfo {author}
  {\bibfnamefont {Alfredo}\ \bibnamefont {Pasquarello}}, \bibinfo {author}
  {\bibfnamefont {Lorenzo}\ \bibnamefont {Paulatto}}, \bibinfo {author}
  {\bibfnamefont {Carlo}\ \bibnamefont {Sbraccia}}, \bibinfo {author}
  {\bibfnamefont {Sandro}\ \bibnamefont {Scandolo}}, \bibinfo {author}
  {\bibfnamefont {Gabriele}\ \bibnamefont {Sclauzero}}, \bibinfo {author}
  {\bibfnamefont {Ari~P}\ \bibnamefont {Seitsonen}}, \bibinfo {author}
  {\bibfnamefont {Alexander}\ \bibnamefont {Smogunov}}, \bibinfo {author}
  {\bibfnamefont {Paolo}\ \bibnamefont {Umari}}, \ and\ \bibinfo {author}
  {\bibfnamefont {Renata~M}\ \bibnamefont {Wentzcovitch}},\ }\bibfield  {title}
  {\enquote {\bibinfo {title} {{QUANTUM} {ESPRESSO}: a modular and open-source
  software project for quantum simulations of materials},}\ }\href {\doibase
  10.1088/0953-8984/21/39/395502} {\bibfield  {journal} {\bibinfo  {journal}
  {Journal of Physics: Condensed Matter}\ }\textbf {\bibinfo {volume} {21}},\
  \bibinfo {pages} {395502} (\bibinfo {year} {2009})}\BibitemShut {NoStop}%
\bibitem [{\citenamefont {Kresse}\ and\ \citenamefont
  {Furthmuller}({1996})}]{Kresse1996}%
  \BibitemOpen
  \bibfield  {author} {\bibinfo {author} {\bibfnamefont {G}~\bibnamefont
  {Kresse}}\ and\ \bibinfo {author} {\bibfnamefont {J}~\bibnamefont
  {Furthmuller}},\ }\bibfield  {title} {\enquote {\bibinfo {title} {{Efficient
  iterative schemes for ab initio total-energy calculations using a plane-wave
  basis set}},}\ }\href {\doibase {10.1103/PhysRevB.54.11169}} {\bibfield
  {journal} {\bibinfo  {journal} {{PHYSICAL REVIEW B}}\ }\textbf {\bibinfo
  {volume} {{54}}},\ \bibinfo {pages} {{11169--11186}} (\bibinfo {year}
  {{1996}})}\BibitemShut {NoStop}%
\bibitem [{\citenamefont {Mostofi}\ \emph {et~al.}(2014)\citenamefont
  {Mostofi}, \citenamefont {Yates}, \citenamefont {Pizzi}, \citenamefont {Lee},
  \citenamefont {Souza}, \citenamefont {Vanderbilt},\ and\ \citenamefont
  {Marzari}}]{Mostofi2014}%
  \BibitemOpen
  \bibfield  {author} {\bibinfo {author} {\bibfnamefont {Arash~A.}\
  \bibnamefont {Mostofi}}, \bibinfo {author} {\bibfnamefont {Jonathan~R.}\
  \bibnamefont {Yates}}, \bibinfo {author} {\bibfnamefont {Giovanni}\
  \bibnamefont {Pizzi}}, \bibinfo {author} {\bibfnamefont {Young-Su}\
  \bibnamefont {Lee}}, \bibinfo {author} {\bibfnamefont {Ivo}\ \bibnamefont
  {Souza}}, \bibinfo {author} {\bibfnamefont {David}\ \bibnamefont
  {Vanderbilt}}, \ and\ \bibinfo {author} {\bibfnamefont {Nicola}\ \bibnamefont
  {Marzari}},\ }\bibfield  {title} {\enquote {\bibinfo {title} {An updated
  version of wannier90: A tool for obtaining maximally-localised wannier
  functions},}\ }\href {\doibase 10.1016/j.cpc.2014.05.003} {\bibfield
  {journal} {\bibinfo  {journal} {Computer Physics Communications}\ }\textbf
  {\bibinfo {volume} {185}},\ \bibinfo {pages} {2309--2310} (\bibinfo {year}
  {2014})}\BibitemShut {NoStop}%
\bibitem [{\citenamefont {Noffsinger}\ \emph {et~al.}(2010)\citenamefont
  {Noffsinger}, \citenamefont {Giustino}, \citenamefont {Malone}, \citenamefont
  {Park}, \citenamefont {Louie},\ and\ \citenamefont {Cohen}}]{Noffsinger2010}%
  \BibitemOpen
  \bibfield  {author} {\bibinfo {author} {\bibfnamefont {Jesse}\ \bibnamefont
  {Noffsinger}}, \bibinfo {author} {\bibfnamefont {Feliciano}\ \bibnamefont
  {Giustino}}, \bibinfo {author} {\bibfnamefont {Brad~D.}\ \bibnamefont
  {Malone}}, \bibinfo {author} {\bibfnamefont {Cheol-Hwan}\ \bibnamefont
  {Park}}, \bibinfo {author} {\bibfnamefont {Steven~G.}\ \bibnamefont {Louie}},
  \ and\ \bibinfo {author} {\bibfnamefont {Marvin~L.}\ \bibnamefont {Cohen}},\
  }\bibfield  {title} {\enquote {\bibinfo {title} {{EPW}: A program for
  calculating the electron{\textendash}phonon coupling using maximally
  localized wannier functions},}\ }\href {\doibase 10.1016/j.cpc.2010.08.027}
  {\bibfield  {journal} {\bibinfo  {journal} {Computer Physics Communications}\
  }\textbf {\bibinfo {volume} {181}},\ \bibinfo {pages} {2140--2148} (\bibinfo
  {year} {2010})}\BibitemShut {NoStop}%
\bibitem [{\citenamefont {Ponc{\'{e}}}\ \emph {et~al.}(2016)\citenamefont
  {Ponc{\'{e}}}, \citenamefont {Margine}, \citenamefont {Verdi},\ and\
  \citenamefont {Giustino}}]{Ponce2016}%
  \BibitemOpen
  \bibfield  {author} {\bibinfo {author} {\bibfnamefont {S.}~\bibnamefont
  {Ponc{\'{e}}}}, \bibinfo {author} {\bibfnamefont {E.R.}\ \bibnamefont
  {Margine}}, \bibinfo {author} {\bibfnamefont {C.}~\bibnamefont {Verdi}}, \
  and\ \bibinfo {author} {\bibfnamefont {F.}~\bibnamefont {Giustino}},\
  }\bibfield  {title} {\enquote {\bibinfo {title} {{EPW}:
  Electron{\textendash}phonon coupling, transport and superconducting
  properties using maximally localized wannier functions},}\ }\href {\doibase
  10.1016/j.cpc.2016.07.028} {\bibfield  {journal} {\bibinfo  {journal}
  {Computer Physics Communications}\ }\textbf {\bibinfo {volume} {209}},\
  \bibinfo {pages} {116--133} (\bibinfo {year} {2016})}\BibitemShut {NoStop}%
\bibitem [{\citenamefont {Xi}\ \emph {et~al.}(2014)\citenamefont {Xi},
  \citenamefont {Wang}, \citenamefont {Yi},\ and\ \citenamefont
  {Shuai}}]{Xi2014}%
  \BibitemOpen
  \bibfield  {author} {\bibinfo {author} {\bibfnamefont {Jinyang}\ \bibnamefont
  {Xi}}, \bibinfo {author} {\bibfnamefont {Dong}\ \bibnamefont {Wang}},
  \bibinfo {author} {\bibfnamefont {Yuanping}\ \bibnamefont {Yi}}, \ and\
  \bibinfo {author} {\bibfnamefont {Zhigang}\ \bibnamefont {Shuai}},\
  }\bibfield  {title} {\enquote {\bibinfo {title} {Electron-phonon couplings
  and carrier mobility in graphynes sheet calculated using the
  wannier-interpolation approach},}\ }\href {\doibase 10.1063/1.4887538}
  {\bibfield  {journal} {\bibinfo  {journal} {The Journal of Chemical Physics}\
  }\textbf {\bibinfo {volume} {141}},\ \bibinfo {pages} {034704} (\bibinfo
  {year} {2014})}\BibitemShut {NoStop}%
\bibitem [{\citenamefont {Baroni}\ \emph {et~al.}(2001)\citenamefont {Baroni},
  \citenamefont {de~Gironcoli}, \citenamefont {Corso},\ and\ \citenamefont
  {Giannozzi}}]{Baroni2001}%
  \BibitemOpen
  \bibfield  {author} {\bibinfo {author} {\bibfnamefont {Stefano}\ \bibnamefont
  {Baroni}}, \bibinfo {author} {\bibfnamefont {Stefano}\ \bibnamefont
  {de~Gironcoli}}, \bibinfo {author} {\bibfnamefont {Andrea~Dal}\ \bibnamefont
  {Corso}}, \ and\ \bibinfo {author} {\bibfnamefont {Paolo}\ \bibnamefont
  {Giannozzi}},\ }\bibfield  {title} {\enquote {\bibinfo {title} {Phonons and
  related crystal properties from density-functional perturbation theory},}\
  }\href {\doibase 10.1103/revmodphys.73.515} {\bibfield  {journal} {\bibinfo
  {journal} {Reviews of Modern Physics}\ }\textbf {\bibinfo {volume} {73}},\
  \bibinfo {pages} {515--562} (\bibinfo {year} {2001})}\BibitemShut {NoStop}%
\bibitem [{\citenamefont {Madsen}\ \emph {et~al.}({2018})\citenamefont
  {Madsen}, \citenamefont {Carrete},\ and\ \citenamefont
  {Verstraete}}]{Madsen2018}%
  \BibitemOpen
  \bibfield  {author} {\bibinfo {author} {\bibfnamefont {Georg K.~H.}\
  \bibnamefont {Madsen}}, \bibinfo {author} {\bibfnamefont {Jesus}\
  \bibnamefont {Carrete}}, \ and\ \bibinfo {author} {\bibfnamefont
  {Matthieu~J.}\ \bibnamefont {Verstraete}},\ }\bibfield  {title} {\enquote
  {\bibinfo {title} {{BoltzTraP2, a program for interpolating band structures
  and calculating semi-classical transport coefficients}},}\ }\href {\doibase
  {10.1016/j.cpc.2018.05.010}} {\bibfield  {journal} {\bibinfo  {journal}
  {{COMPUTER PHYSICS COMMUNICATIONS}}\ }\textbf {\bibinfo {volume} {{231}}},\
  \bibinfo {pages} {{140--145}} (\bibinfo {year} {{2018}})}\BibitemShut
  {NoStop}%
\bibitem [{\citenamefont {Madsen}\ and\ \citenamefont
  {Singh}(2006)}]{Madsen2006}%
  \BibitemOpen
  \bibfield  {author} {\bibinfo {author} {\bibfnamefont {Georg~K.H.}\
  \bibnamefont {Madsen}}\ and\ \bibinfo {author} {\bibfnamefont {David~J.}\
  \bibnamefont {Singh}},\ }\bibfield  {title} {\enquote {\bibinfo {title}
  {Boltztrap. a code for calculating band-structure dependent quantities},}\
  }\href {\doibase 10.1016/j.cpc.2006.03.007} {\bibfield  {journal} {\bibinfo
  {journal} {Computer Physics Communications}\ }\textbf {\bibinfo {volume}
  {175}},\ \bibinfo {pages} {67--71} (\bibinfo {year} {2006})}\BibitemShut
  {NoStop}%
\bibitem [{\citenamefont {Yang}\ \emph {et~al.}(2008)\citenamefont {Yang},
  \citenamefont {Li}, \citenamefont {Wu}, \citenamefont {Zhang}, \citenamefont
  {Chen},\ and\ \citenamefont {Yang}}]{Yang2008}%
  \BibitemOpen
  \bibfield  {author} {\bibinfo {author} {\bibfnamefont {Jiong}\ \bibnamefont
  {Yang}}, \bibinfo {author} {\bibfnamefont {Huanming}\ \bibnamefont {Li}},
  \bibinfo {author} {\bibfnamefont {Ting}\ \bibnamefont {Wu}}, \bibinfo
  {author} {\bibfnamefont {Wenqing}\ \bibnamefont {Zhang}}, \bibinfo {author}
  {\bibfnamefont {Lidong}\ \bibnamefont {Chen}}, \ and\ \bibinfo {author}
  {\bibfnamefont {Jihui}\ \bibnamefont {Yang}},\ }\bibfield  {title} {\enquote
  {\bibinfo {title} {Evaluation of half-heusler compounds as thermoelectric
  materials based on the calculated electrical transport properties},}\ }\href
  {\doibase 10.1002/adfm.200701369} {\bibfield  {journal} {\bibinfo  {journal}
  {Advanced Functional Materials}\ }\textbf {\bibinfo {volume} {18}},\ \bibinfo
  {pages} {2880--2888} (\bibinfo {year} {2008})}\BibitemShut {NoStop}%
\bibitem [{\citenamefont {Hong}\ \emph {et~al.}(2016)\citenamefont {Hong},
  \citenamefont {Gong}, \citenamefont {Li}, \citenamefont {Yan}, \citenamefont
  {Ren},\ and\ \citenamefont {Liu}}]{Hong2016}%
  \BibitemOpen
  \bibfield  {author} {\bibinfo {author} {\bibfnamefont {A.~J.}\ \bibnamefont
  {Hong}}, \bibinfo {author} {\bibfnamefont {J.~J.}\ \bibnamefont {Gong}},
  \bibinfo {author} {\bibfnamefont {L.}~\bibnamefont {Li}}, \bibinfo {author}
  {\bibfnamefont {Z.~B.}\ \bibnamefont {Yan}}, \bibinfo {author} {\bibfnamefont
  {Z.~F.}\ \bibnamefont {Ren}}, \ and\ \bibinfo {author} {\bibfnamefont
  {J.-M.}\ \bibnamefont {Liu}},\ }\bibfield  {title} {\enquote {\bibinfo
  {title} {Predicting high thermoelectric performance of abx ternary compounds
  namgx (x = p, sb, as) with weak electron-phonon coupling and strong bonding
  anharmonicity},}\ }\href {\doibase 10.1039/C6TC00461J} {\bibfield  {journal}
  {\bibinfo  {journal} {J. Mater. Chem. C}\ }\textbf {\bibinfo {volume} {4}},\
  \bibinfo {pages} {3281--3289} (\bibinfo {year} {2016})}\BibitemShut {NoStop}%
\end{thebibliography}
%

\end{document}